\documentclass[a4paper,11pt]{article} 
\pdfoutput=1
\usepackage{jheppub} 
\hypersetup{unicode=true}
\usepackage[utf8x]{inputenc} 
\usepackage[T1]{fontenc} 
\usepackage{bbm,bm,commath,mathtools,wasysym,xfrac} 
\usepackage[nottoc,notlof,notlot]{tocbibind} 

\usepackage{xcolor} 
\usepackage{trimclip} 
\usepackage{stackengine} 
\usepackage{tikz} 
\usepackage{tikz-cd} 
\usepackage{tikz-3dplot} 
\graphicspath{{figures/}} 

\DeclareUnicodeCharacter{"0393}{\Gamma}
\DeclareUnicodeCharacter{"0394}{\Delta}
\DeclareUnicodeCharacter{"0398}{\Theta}
\DeclareUnicodeCharacter{"039B}{\Lambda}
\DeclareUnicodeCharacter{"039E}{\Xi}
\DeclareUnicodeCharacter{"03A0}{\Pi}
\DeclareUnicodeCharacter{"03A3}{\Sigma}
\DeclareUnicodeCharacter{"03A5}{\Upsilon}
\DeclareUnicodeCharacter{"03A6}{\Phi}
\DeclareUnicodeCharacter{"03A8}{\Psi}
\DeclareUnicodeCharacter{"03A9}{\Omega}
\DeclareUnicodeCharacter{"03B1}{\alpha}
\DeclareUnicodeCharacter{"03B2}{\beta}
\DeclareUnicodeCharacter{"03B3}{\gamma}
\DeclareUnicodeCharacter{"03B4}{\delta}
\DeclareUnicodeCharacter{"03B5}{\epsilon}
\DeclareUnicodeCharacter{"03B6}{\zeta}
\DeclareUnicodeCharacter{"03B7}{\eta}
\DeclareUnicodeCharacter{"03B8}{\theta}
\DeclareUnicodeCharacter{"03D1}{\vartheta}
\DeclareUnicodeCharacter{"03B9}{\iota}
\DeclareUnicodeCharacter{"03BA}{\kappa}
\DeclareUnicodeCharacter{"03BB}{\lambda}
\DeclareUnicodeCharacter{"03BC}{\mu}
\DeclareUnicodeCharacter{"03BD}{\nu}
\DeclareUnicodeCharacter{"03BE}{\xi}
\DeclareUnicodeCharacter{"03C0}{\pi}
\DeclareUnicodeCharacter{"03C1}{\rho}
\DeclareUnicodeCharacter{"03C3}{\sigma} 
\DeclareUnicodeCharacter{"03C4}{\tau}
\DeclareUnicodeCharacter{"03C5}{\upsilon}
\DeclareUnicodeCharacter{"03C6}{\phi}
\DeclareUnicodeCharacter{"03D5}{\varphi}
\DeclareUnicodeCharacter{"03C7}{\chi}
\DeclareUnicodeCharacter{"03C8}{\psi}
\DeclareUnicodeCharacter{"03C9}{\omega}
\DeclareUnicodeCharacter{"21D0}{\Leftarrow}
\DeclareUnicodeCharacter{"0212B}{\AA}
\DeclareUnicodeCharacter{"00B7}{\cdot}
\DeclareUnicodeCharacter{"00B0}{^{\circ}}
\DeclareUnicodeCharacter{"266A}{\eighthnote}
\DeclareUnicodeCharacter{"266B}{\twonotes}
\PrerenderUnicode{¹}\PrerenderUnicode{²}
\PrerenderUnicode{₃}\PrerenderUnicode{×}

\newcommand\eqs[1] {\begin{align}#1\end{align}}
\newcommand\eqsn[1] {\begin{align*}#1\end{align*}}
\newcommand\eqst[1] {\begin{multline}#1\end{multline}}

\newcommand\eqsc[1] {\begin{gather}#1\end{gather}}

\newcommand\eqsa[1] {\equ{\begin{aligned}#1\end{aligned}}}
\newcommand\eqsg[1] {\equ{\begin{gathered}#1\end{gathered}}}
\newcommand\equ[1] {\begin{equation}#1\end{equation}}

\newcommand\fig[2] {\begin{figure}[#1]\centering #2\end{figure}}
\newcommand\pmat[1] {\begin{pmatrix}#1\end{pmatrix}}

\newcommand\half {\tfrac{1}{2}}
\newcommand\s {\sigma}

\renewcommand\( {\left(}
\renewcommand\) {\right)}
\newcommand\wh {\widehat}
\newcommand\wt {\widetilde}
\renewcommand\Re {\text{Re}}

\DeclareMathOperator{\Li}{Li}

\def\C {{\mathcal C}}
\newcommand\D {{\mathcal D}}

\renewcommand\H {{\mathcal H}}

\newcommand\J {{\mathcal J}}

\renewcommand\L {{\mathcal L}}
\newcommand\M {{\mathcal M}}
\newcommand\N {{\mathcal N}} 

\newcommand\R {{\mathcal R}}

\newcommand\T {{\mathcal T}}
\def\U {{\mathcal U}}
\newcommand\V {{\mathcal V}}

\newcommand\bA {{\mathbb A}}
\newcommand\bB {{\mathbb B}}
\newcommand\bC {{\mathbb C}}

\newcommand\bP {{\mathbb P}}
\newcommand\bR {{\mathbb R}}

\newcommand\bZ {{\mathbb Z}}

\newcommand\sT {{\sf T}} 

\definecolor{darkgreen}{rgb}{0,0.5,0}

\newcommand\nn {\nonumber\\}
\newcommand\cpone {\bC\bP^1}
\newcommand\cptwo {\bC\bP^2}
\newcommand\cpnnn {\bC\bP^N}

\newcommand\vc[1] {\vcenter{\hbox{#1}}}

\newcommand\twoclr[4]{%
  \mbox{%
    \def\Whalf##1{\dimexpr.5\width ##1 #4em\relax}
    \textcolor{#1}{\clipbox{0em -0.1em {\Whalf+} -0.1em}{$#3$}}%
    \textcolor{#2}{\clipbox{{\Whalf-} -0.1em 0em -0.1em}{$#3$}}%
  }%
}
\newcommand\forclr[7]{%
  \mbox{%
    \def\Whalf##1{\dimexpr.5\width ##1 #6em\relax}
    \def\Hhalf##1{\dimexpr.5\totalheight ##1 #7em\relax}
    \stackanchor[0pt]{\textcolor{#1}{\clipbox{0em {\Hhalf-} {\Whalf+} -0.1em}{$#5$}}%
    \textcolor{#2}{\clipbox{{\Whalf-} {\Hhalf-} 0em -0.1em}{$#5$}}}%
    {\textcolor{#4}{\clipbox{0em -0.1em {\Whalf+} {\Hhalf+}}{$#5$}}%
    \textcolor{#3}{\clipbox{{\Whalf-} -0.1em 0em {\Hhalf+}}{$#5$}}}%
  }%
}

\numberwithin{equation}{section} 
\makeatletter
\gdef\@fpheader{Journal version: \href{https://dx.doi.org/10.1007/JHEP09(2021)112}{{\it JHEP} {\bf 09} (2021) 112}}
\makeatother

\begin{document}
\title{\texorpdfstring{Stokes Phenomena in 3d $\N\bm{=2}$ SQED$\bm{{}_2}$ and $\bm{\cpone}$ Models}{Stokes Phenomena in 3d N=2 SQED₂ and CP¹ Models}}

\author[ψ]{Dharmesh Jain\href{https://orcid.org/0000-0002-9310-7012}{\includegraphics[scale=0.0775]{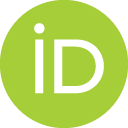}}}
\affiliation[ψ]{Department of Theoretical Sciences,\\
S. N. Bose National Centre for Basic Sciences,\\
Block--JD, Sector--III, Salt Lake City,\\
Kolkata 700106, India\\}
\emailAdd{dharmesh.jain@bose.res.in}

\author[μ]{and Arkajyoti Manna\href{https://orcid.org/0000-0002-4772-1150}{\includegraphics[scale=0.0775]{ORCIDiD_icon128x128.png}}}
\affiliation[μ]{Institute of Mathematical Sciences,\\
Homi Bhabha National Institute (HBNI),\\
IV Cross Road, C.I.T. Campus, Taramani,\\
Chennai 600113, India\\}
\emailAdd{arkajyotim@imsc.res.in}

\abstract{We propose a novel approach of uncovering Stokes phenomenon exhibited by the holomorphic blocks of $\cpone$ model by considering it as a specific decoupling limit of SQED${}_2$ model. This approach involves using a $\bZ_3$ symmetry that leaves the supersymmetric parameter space of SQED${}_2$ model invariant to transform a pair of SQED${}_2$ holomorphic blocks to get two new pairs of blocks. The original pair obtained by solving the line operator identities of the SQED${}_2$ model and the two new transformed pairs turn out to be related by Stokes-like matrices. These three pairs of holomorphic blocks can be reduced to the known triplet of $\cpone$ blocks in a particular decoupling limit where two of the chiral multiplets in the SQED${}_2$ model are made infinitely massive. This reduction then correctly reproduces the Stokes regions and matrices of the $\cpone$ blocks. Along the way, we find six pairs of SQED${}_2$ holomorphic blocks in total, which lead to six Stokes-like regions covering uniquely the full parameter space of the SQED${}_2$ model.
}


\maketitle

\section{Introduction}
The three-dimensional supersymmetric field theories exhibit interesting dynamics and intriguing dualities \cite{Intriligator:1996ex,Aharony:1997bx,Dorey:1999rb,Tong:2000ky}. Their study got a new direction with the development of supersymmetric localization techniques \cite{Witten:1988ze,Pestun:2007rz}. Supersymmetric partition functions on various curved manifolds can be computed exactly, allowing one to study explicit examples of RG flows nonperturbatively, and providing nontrivial checks of various dualities including holography \cite{Kapustin:2009kz,Jafferis:2010un,Hama:2010av,Herzog:2010hf,Hama:2011ea,Imamura:2011wg,Nian:2013qwa,Imamura:2013qxa}. With the systematic study of rigid supersymmetry on curved manifolds \cite{Festuccia:2011ws,Closset:2012ru}, all possible manifolds on which at least $\N=2$ supersymmetry can be preserved have been classified and corresponding partition functions $Z_{M_3}$ have been computed, with the possible exception of the 3-torus $T^3$ \cite{Closset:2017zgf,Closset:2018ghr,Pittelli:2018rpl,Closset:2019hyt}. Among them is the partition function on a manifold given by the twisted product of a disk (2d hemisphere) and a circle: $D^2\,×_q\,S^1$ \cite{Yoshida:2014ssa,Bullimore:2020jdq,Crew:2020psc}, where $q$ is a deformation parameter of $D^2$. This partition function is also known as a \emph{holomorphic block}.

The holomorphic blocks were discovered originally when other partition functions like $Z_{S^3}$, or the 3d (twisted) superconformal index $Z_{S^2×S^1}$ factorized into products of common ``building blocks'' \cite{Pasquetti:2011fj,Dimofte:2011py,Beem:2012mb}.\footnote{The holomorphic blocks also exist in 4d and 5d theories where factorization of partition functions is again observed \cite{Yoshida:2014qwa,Nieri:2015yia,Pasquetti:2016dyl}. The 4d holomorphic blocks were later derived directly by localization on $D^2×T^2$ in \cite{Longhi:2019hdh}, which also includes an analysis of boundary conditions on $T^3$.} These blocks were quickly realized to have an independent existence of their own as $Z_{D^2×_q S^1}$ with more intricate analytical properties, which are not at all obvious from the partition functions obtained after gluing the blocks but are crucial for the gluing procedure to work. For example, there can be various distinct sets of these blocks, which when glued appropriately, produce the same partition function. The study of these analytical properties has not been a major focus in the literature and so we are going to make it the subject of this note.

Before we dive into these analytical properties, let us discuss the physical motivation for studying the holomorphic block $(\bB^α)$, which is in one-to-one correspondence with the vacuum $α$ for the 3d massive theory compactified on a circle. It can thus be interpreted as a ``BPS index'' counting BPS states in such a vacuum of the massive theory obtained by deforming a superconformal field theory (SCFT) by relevant operators \cite{Beem:2012mb}. Such a massive deformation of generic $\N=2$ SCFTs is obtained by turning on real masses ($m_i$) corresponding to any $U(1)$ subgroup in the flavour symmetry group of the theory. For topological $U(1)$ symmetries, these real masses appear as Fayet-Iliopoulos (FI) parameters $(ζ_a)$. The holomorphic blocks are then expected to be analytical functions in the space of these mass and FI parameters. Typically, it is found that upon varying these parameters, the holomorphic blocks are subjected to Stokes phenomenon, i.e., the blocks $\bB^α$'s associated to vacua $α$ in one region of parameter space get connected to blocks in a different region by a linear transformation:
\equ{\bB^α \longrightarrow {\bB'}^α=∑_β M^α{}_β \bB^β\,, \qquad M^α{}_β∈GL(N,\bZ)\,,
\label{defSMs}}
where $M^α{}_β$ is called the connection or Stokes matrix. This behaviour is physically expected since vacuum $α$ may change as mass parameters are varied.\footnote{The relevant variable needed for expanding a given holomorphic block as a $q$-series may be a specific combination of the various parameters appearing in the theory under consideration, which could also differ from region to region in the parameter space. Thus, Stokes phenomenon can be thought of as a consequence of noncommutativity of series expansion and analytic continuation. See \cite{Kawai:2005yt} for the typical example of Stokes behaviour exhibited by Airy function.} Thus the study of holomorphic blocks and the associated Stokes behaviour gives physical insights about the BPS states and how their counting changes across the parameter space. 

An exploratory study of the analytical properties of the holomorphic blocks in the particular case of $\cpone$ model \cite{Dorey:1999rb,Tong:2000ky} has been undertaken in \cite{Beem:2012mb}. It is based on a prescriptive approach of constructing certain block integrands whose specified contour integrals give rise to the holomorphic blocks. The deformation of the associated contours then gives rise to Stokes phenomenon in the parameter space spanned by (complexified) masses and FI parameters.\footnote{The exact result for partition functions like $Z_{S^3}$ when cast as a resurgent transseries can exhibit a different kind of Stokes phenomenon as studied in \cite{Fujimori:2018nvz,Fujimori:2021oqg}.} For $\cpone$ model, this phenomenon was shown to be related to the mirror symmetry. But it is more general, as shown in \cite{Ashok:2019gee} by following an algebraic approach of constructing the holomorphic blocks. This algebraic approach involves solving certain $q$-difference equations called line operator identities (LOIs) which are known to annihilate the holomorphic blocks as discussed in \cite{Beem:2012mb}. The Stokes phenomenon then becomes a direct consequence of the singularity structure of the $q$-difference equations in question. For $\cpone$ model, the irregular singularity (see \cite{TAB} for example, for the definition of (ir)regular singular points) of the LOIs satisfied by the holomorphic blocks (given in terms of the $q$-hypergeometric function, ${}_1\phi_1(0;a;q,z)$ \cite{OHY2016}) leads to the expected Stokes matrices found via the block integral approach.

The algebraic approach for constructing holomorphic blocks is best suited to generalization to $\cpnnn$ models \cite{Jockers:2018sfl} since it involves no guesswork regarding the choice or deformation of contours, a procedure which becomes quite unwieldy very quickly even for $N=2$. In fact, half of the connection formulae required for the study of Stokes phenomenon in $\cptwo$ are (more or less) available in the mathematical literature (see \cite{OHY2017,OHYtalk2} and references therein) and the other half that requires taming the irregular singularity is the only technical bottleneck in understanding the full analytical structure of these holomorphic blocks. Similar situation occurs for $N>2$ and the technicalities are only expected to get worse. So we propose an alternative approach in this note to circumvent the technical issues associated with the irregular singularity of the LOIs associated with the $\cpnnn$ models.

Our proposal is to extract $\cpnnn$ model's holomorphic blocks by taking a suitable decoupling limit of the blocks of SQED${}_{N+1}$ model \cite{Intriligator:1996ex,deBoer:1997ka,Nieri:2015yia}. The SQED${}_{N+1}$ models have twice the number of chiral multiplets when compared to the $\cpnnn$ models, so we decouple a half of the multiplets by making them infinitely massive in a specific manner. The first step of our proposal is then to solve the LOIs for the SQED models to get two sets of holomorphic blocks, which is relatively easy as these LOIs have two regular singularities. Naively, there is no Stokes phenomenon in these models due to the regular behaviour of these singularities and the decoupling limit is unlikely to produce all possible $\cpnnn$ holomorphic blocks that exhibit Stokes phenomenon. This is because even in $\cpone$, there are three independent sets of holomorphic blocks. Thus, to get the nontrivial Stokes phenomenon exhibited by $\cpnnn$ blocks, we have to somehow inject in the ``irregular'' behaviour of one of the singularity of the $\cpnnn$ LOIs. We conjecture that this can be achieved via transformation of the SQED holomorphic blocks (obtained in the first step around regular singularities) by certain discrete transformations that leave invariant the supersymmetric parameter space $\L_{SUSY}$ of the SQED model compactified on $S^1$. This second step is reminiscent of the mirror symmetry transformations acting on $\cpone$ blocks which is related to Stokes phenomenon \cite{Beem:2012mb,Ashok:2019gee}.

These discrete transformations for SQED model do have some flavour of the mirror symmetry as they also transform Coulomb and Higgs branches but we are not invoking a dual theory for the description of the theory after such transformations. We remain in the original SQED model to study how the holomorphic blocks behave in the same parameter space and take the decoupling limit of the transformed holomorphic blocks. Once all the ingredients fit into place, we expect this limit to reproduce the Stokes phenomenon expected from the holomorphic blocks of $\cpnnn$ model. At this point, one might object that we seem to have traded the guesswork attributed to the choice of contours in the block integral approach with the one involved in finding the transformations that leave $\L_{SUSY}$ of the SQED model invariant. However, combining the knowledge of the explicit form of $\L_{SUSY}$ in terms of FI and flavour parameters, and appearance of only certain combinations of these parameters in the SQED holomorphic blocks, it is quite easy to read off certain discrete transformations. The only caveat being whether these transformations are relevant for our purposes of reducing the SQED model via a well-defined decoupling limit to the corresponding $\bC\bP$ model. Instead of tackling this question in general, we will specialize to the case of SQED${}_2$ model in this note as a proof of concept. We will show that a $\bZ_3$ symmetry which leaves $\L_{SUSY}$ invariant readily provides the required transformations for the SQED${}_2$ model, which indeed allow us to recover Stokes matrices of the $\cpone$ model in a specific decoupling limit. In addition, this $\bZ_3$ symmetry leads to six pairs of holomorphic blocks for SQED${}_2$ model, whose regions of validity uniquely cover the whole of the model's parameter space, providing the Stokes-like\footnote{We use the phrase ``Stokes-like'' in the case of SQED${}_2$ model since not all connection matrices belong to $GL(N,\bZ)$, with some matrix entries being elliptic factors instead of integers.} partitioning of the full parameter space.

The rest of this note is organized as follows. In Section \ref{sec:SQED2model}, we study the 3d SQED${}_2$ model with an emphasis on its phase diagram and $\L_{SUSY}$ obtained by compactification on $S^1$. A $\bZ_3$ symmetry of the latter provides us two transformations that will be used to demonstrate that the proposal outlined above can be successfully implemented. Next, we solve the SQED${}_2$ LOIs to get two pairs of holomorphic blocks in Section \ref{sec:HoloBlocks}. In Section \ref{sec:DiscreteTrans}, we will transform these pairs of blocks using the two transformations forming the $\bZ_3$ group (along with the identity transformation) to get new sets of holomorphic blocks for the SQED${}_2$ model. This will provide us with a unique partitioning of the full parameter space of the SQED${}_2$ model into six Stokes-like regions. Finally, in Section \ref{sec:DecLim}, we identify the correct decoupling limit that reproduces the three pairs of $\cpone$ holomorphic blocks from a particular triplet of SQED${}_2$ ones along with the correct Stokes regions and matrices. We conclude with future outlook in Section \ref{sec:Outlook} and three technical Appendices \ref{app:FunDefIds}, \ref{app:DeriveHBs} and \ref{app:ExtraDTs}.

\section{\texorpdfstring{SQED$\bm{{}_2}$ Model}{SQED₂ Model}}\label{sec:SQED2model}
We start by setting up the 3d $\N=2$ SQED${}_2$ model \cite{Intriligator:1996ex,deBoer:1997ka}. This is a $U(1)$ gauge theory with 4 chiral fields $φ_{1,2,3,4}$ and three flavour symmetries $U(1)^2×U(1)_J.$\footnote{There is one more axial $U(1)$ symmetry under which all the four fields have unit charge. We do not turn on a mass for this symmetry because it breaks CP invariance \cite{Aharony:1997bx}.} The charge matrix and Chern-Simons (CS) coefficient matrix are given by:\footnote{\label{fn2}We assign a nonzero R-charge to one of the chiral fields for technical reasons. This ensures that the SQED${}_2$ blocks we derive later reduce to $\cpone$ blocks exactly in the $m_2→∞$ limit.}
\begin{equation}
T^{SQED_2}[\vec{φ}\,]=\left\{\vphantom{\pmat{ \\[2.2cm] }}\right.\quad
\begin{array}{c|cccc}Q & φ_1 & φ_2 & φ_3 & φ_4 \\ \hline
G & 1 & 1 & -1 & -1 \\
F_1 & 1 & -1 & 0 & 0 \\
F_2 & 0 & 0 & 1 & -1 \\
J & 0 & 0 & 0 & 0 \\
R & 0 & 0 & 0 & 2
\end{array} \qquad
\begin{array}{c|ccccc}k & G & F_1 & F_2 & J & R \\ \hline
G & 0 & 0 & 0 & 1 & 0 \\
F_1 & 0 & 0 & 0 & 0 & 0 \\
F_2 & 0 & 0 & 0 & 0 & 0\\
J & 1 & 0 & 0 & 0 & 0 \\
R & 0 & 0 & 0 & 0 & ⋆
\end{array}
\label{sqed2model}
\end{equation}
We can turn on real masses $(m_1,m_2)$ corresponding to the two flavour $U(1)$'s and a FI parameter $ζ$, which parameterize the space of vacuum solutions of this model. These solutions are obtained by setting the scalar potential of the model to zero. This potential can be constructed quite easily (see \cite{Dorey:1999rb} for example) and it reads:
\eqst{U=e^2(|φ_1|^2+|φ_2|^2-|φ_3|^2-|φ_4|^2-ζ^{\rm{eff}})^2+(\s+m_1)^2|φ_1|^2+(\s-m_1)^2|φ_2|^2 \\
+(-\s+m_2)^2|φ_3|^2+(-\s-m_2)^2|φ_4|^2\,,
}
where $e$ is the $U(1)$ gauge coupling constant (with mass dimension $\frac{1}{2}$, same as that of the complex scalars $φ_i$), $\s$ is the real scalar in the $U(1)$ vector multiplet (with mass dimension $1$) and $ζ^{\rm{eff}}=ζ+\frac{1}{2}|\s+m_1|+\frac{1}{2}|\s-m_1|+\frac{1}{2}|-\s+m_2|+\frac{1}{2}|-\s-m_2|$.

The distribution of vacua in the three-dimensional parameter space $\{m_1,m_2,ζ\}$ is shown in Figure \ref{fig:SQED2PD}. The Higgs branch of vacua is depicted by blue planes that are given by
\equ{{\color{blue}\H=\{m_1,m_2,ζ \,|\, m_1=0\,\cap\,ζ-|m_2|>0\,;\; m_2=0\,\cap\,ζ+|m_1|<0\}}\,,
}
the Coulomb branch by orange planes:
\equ{{\color{orange}\C=\{m_1,m_2,ζ \,|\, ζ=0\,;\; ζ+|m_1|-|m_2|=0\}}\,,
}
and the mixed branch by green planes:\footnote{We thank M. Martone for showing off his coding abilities in \cite{Martone:2021ixp}.}
\equ{\twoclr{blue}{orange}{\mathcal{M}}{0.025}={\color{darkgreen}\{m_1,m_2,ζ \,|\, |m_1|-|m_2|=0\}}\,.
}
The SQED${}_2$ model has two massive Coulomb vacua in between the orange planes, and the rest of the parameter space admits two massive Higgs/mixed vacua.
\fig{!h}{\includegraphics[scale=0.55]{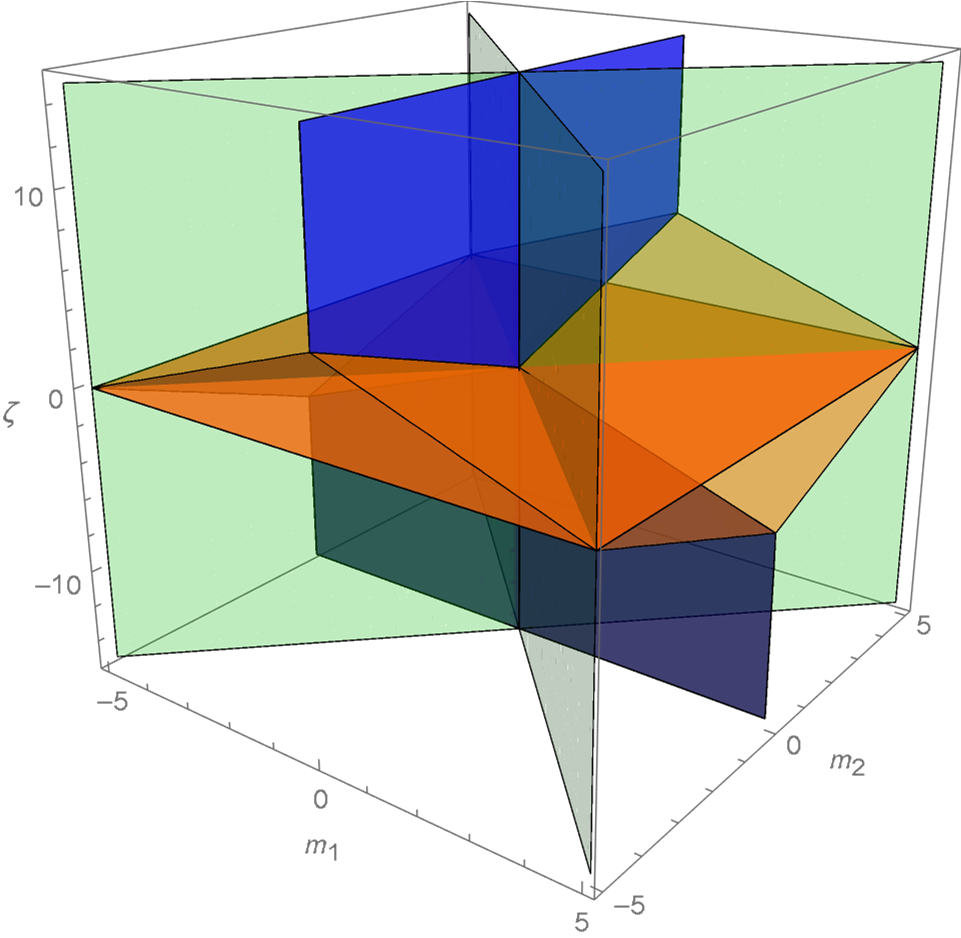}
\caption{Phase diagram of the SQED${}_2$ model. The blue, orange, and green planes denote Higgs, Coulomb, and mixed branches, respectively. The remaining parameter space admits only two massive vacuum solutions.}
\label{fig:SQED2PD}
}

\paragraph{Compactification on $\bm{S^1}$.} Now let us consider the compactification of the SQED${}_2$ model on a circle. The mass and FI parameters are complexified due to Wilson lines as $m_{1,2}→X_{1,2}$ and $ζ→T$. The single-valued complex variables are then defined via exponentiation as $x_{1,2}=e^{X_{1,2}}$ and $t=e^T$. Additionally, we define the scalar in the 2d vector multiplet as $s=e^S$, which is the complexification of $\s$. The effective twisted superpotential in two dimensions is then given by (up to constants)
\begin{align}
\wt{W}(S;X_{1,2},T) &=\tfrac{1}{4}(S±X_1-iπ)^2+\Li_2\big(e^{-S∓X_1}\big) \nn
&\qquad +\tfrac{1}{4}(-S±X_2\mp iπ)^2+\Li_2\big(e^{S∓X_2}\big) +ST \nn
&=\tfrac{1}{2}(X_1^2+ X_2^2)-i\pi (S+X_2)+S(S+T) \nn
&\qquad +\Li_2(e^{-S∓X_1}) +\Li_2\big(e^{S∓X_2}\big)\,.
\label{Wtsqed2Def}
\end{align}
From the twisted superpotential, we can determine the supersymmetric parameter space by solving the equations
\equ{e^{s∂_{s}\wt{W}}=1\,,\qquad e^{x_i∂_{x_i}\wt{W}}=p_i\,, \quad i=1,2\,.
}
Here, $p_i$ are the effective background FI parameters for the flavour symmetries that allow supersymmetry to be preserved, if the flavour symmetries were ever gauged. Now, eliminating $s$ leads to the following polynomial equations that characterize the susy parameter space:
\equ{\L_{SUSY}=\left\{\begin{aligned}
(1-x_2p_t)(1-x_2^{-1}p_t)+t(1-x_1p_t)(1-x_1^{-1}p_t)&\simeq 0 \\
p_1p_t -x_1 (p_1+p_t)+1 &\simeq 0 \\
p_2p_t -x_2^{-1}(p_2 -p_t)-1&\simeq 0\,.
\end{aligned}\right. 
\label{ClLOIsqed2}}

Next, let us understand the discriminant locus of this model, which maps out the massless vacua. It is obtained when the two solutions for $s$, found by solving $e^{s∂_s\wt{W}}=1$, become degenerate:
\eqsc{s_{±} =\tfrac{x_1(1+x_2^2) +x_2(1+x_1^2)t ±\sqrt{(x_1(1+x_2^2) +x_2(1+x_1^2)t)^2-(2x_1x_2(1+t))^2}}{2x_1x_2(1+t)} \nn
⇒\D:\;\left\{t=-\tfrac{x_2+x_2^{-1}±2}{x_1+x_1^{-1}\pm 2}\right\}·
\label{spmD}}
For real values of these parameters (up to an $iπ$ shift of $X_1$), the discriminant locus is shown in Figure \ref{fig:SQED2PD2}. In the decompactification limit, we recover most of the Figure \ref{fig:SQED2PD}, except the $ζ=0$ and $|m_1|-|m_2|=0$ planes. The vacuum solutions corresponding to these planes that are lost upon compactification have nonzero but unbounded vev for the scalar in the vector multiplet. It seems this reduction in the vacuum solutions compared to the flat $\bR^3$ case is not well-documented in the literature and it would be interesting to have a general understanding of this phenomena.\footnote{While we were finalizing this note, a preprint \cite{Gu:2021} appeared that analyzes this mismatch between the phases of 3d $\N=2$ theories with nonzero superpotential and $m_i=0$. In this case, the mismatch is due to discrete, massive vacua with nonzero vev for the vector multiplet scalar; identified as being in ``topological phase''.}
\fig{!h}{\includegraphics[scale=0.55]{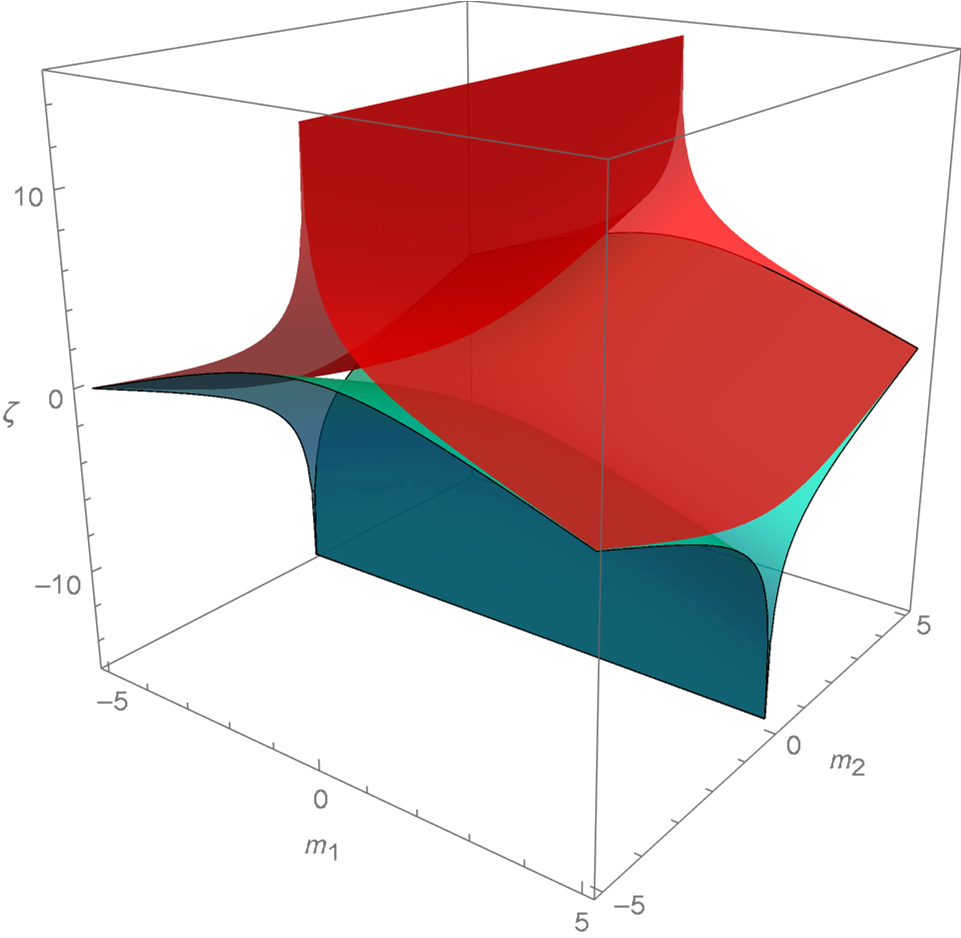}
\caption{Discriminant locus of SQED${}_2$ model compactified on a circle. The red sheet corresponds to the top sign in \eqref{spmD} and cyan sheet to the bottom sign.}
\label{fig:SQED2PD2}}

\paragraph{Discrete Symmetry Transformations.} The discriminant locus as seen in Figure \ref{fig:SQED2PD2} has two obvious $\bZ_2$ symmetries, $\U: m_1→-m_1 ⇒ x_1→x_1^{-1}$ and $\V: m_2→-m_2 ⇒ x_2→x_2^{-1}$. These symmetries will not help us in generating holomorphic blocks that exhibit Stokes phenomenon so we turn to higher-order symmetries. The $m_2→±∞$ (and $m_1→±∞$) boundary of the discriminant locus is reminiscent of the parameter space of $\cpone$ and we might expect a $\bZ_3$ symmetry that leaves the complete discriminant locus invariant. Indeed, we can find such symmetry transformations (the signs are fixed post priori from the holomorphic blocks given in the next section):
\eqsa{\T_{I} &:\quad x_1 \rightarrow \sqrt{e^{i\pi}(x_1x_2)^{-1}t}\,,\quad x_2 \rightarrow \sqrt{e^{i\pi}x_1x_2 t}\,,\quad t \rightarrow -x_1^{-1}x_2\,; \\
(\T_I)^2≡\T_{II} &:\quad x_1 \rightarrow \sqrt{e^{-i\pi}x_1^{-1}x_2t^{-1}}\,,\quad x_2 \rightarrow \sqrt{e^{i\pi}x_1^{-1}x_2 t}\,,\quad t \rightarrow -x_1x_2\,.
\label{DTs1n2}}
The above transformations of the parameter space (with $(\T_I)^3=1$) form the $\mathbb{Z}_3$ symmetry group and also leave the susy parameter space invariant.\footnote{There also exists another $\bZ_3$ symmetry that can be obtained by inverting $t$ in $\T_I$. This is discussed separately in Appendix \ref{app:ExtraDTs}.} The invariance of the susy parameter space $\L_{SUSY}$ given in \eqref{ClLOIsqed2} can be proven by transforming $\wt{W}$ in \eqref{Wtsqed2Def} and introducing appropriate contact terms as follows:
\begingroup
\allowdisplaybreaks
\eqsn{\wt{W}_{\T_I}(S;X_1,X_2,T) &=\wt{W}\(S-\half(X_1+X_2+T+iπ);-\half(X_1+X_2-T-iπ),\right. \nn
&\quad \left.\half(X_1+X_2+T+iπ),-X_1+X_2-iπ\) +\half iπ(X_1-X_2+T)\,; \\
\wt{W}_{\T_{II}}(S;X_1,X_2,T) &=\wt{W}\(S-\half(X_1+X_2+T+iπ);-\half(X_1-X_2+T+iπ),\right. \nn
&\quad \left.\half(-X_1+X_2+T+iπ),X_1+X_2-iπ\) -\half iπ(X_1+X_2-T)\,.
}
\endgroup
The invariance of $\D$ then can be easily checked by solving for transformed $s_±$'s as done in \eqref{spmD} for the original $\wt{W}$.

We end here the general discussion of the SQED${}_2$ model and move on to study the holomorphic blocks of this model now.

\section{Holomorphic Blocks}\label{sec:HoloBlocks}
Following the prescription laid out in \cite{Beem:2012mb}, the line operator identities (LOIs) of SQED${}_2$ model can be found to be
\eqsa{(1-q^{-1}\widehat{x}_2\widehat{p}_t)(1-\widehat{x}_2^{-1}\widehat{p}_t)+\sqrt{q}\,\widehat{t}\,(1-\widehat{x}_1 \widehat{p}_t)(1-\widehat{x}_1^{-1}\widehat{p}_t)&\simeq 0 \\
q^{-\frac{1}{2}}\widehat{p}_1\widehat{p}_t -\widehat{x}_1 (\sqrt{q}\widehat{p}_1+\widehat{p}_t)+1 &\simeq 0 \\
\widehat{p}_2\widehat{p}_t -\widehat{x}_2^{-1}(\widehat{p}_2 -\widehat{p}_t)-1&\simeq 0\,.
\label{LOIsqed2}}
These $q$-difference equations match the equations defining the susy parameter space in \eqref{ClLOIsqed2} in the classical limit $q→1$. We can derive the holomorphic blocks \cite{Nieri:2015yia} as solutions to these difference equations. In the Appendix \ref{app:DeriveHBs}, we present a systematic derivation of one set of holomorphic blocks starting from these LOIs. Here we directly present the solutions and discuss their properties.

The holomorphic blocks near $t=\infty$ are
\begingroup
\allowdisplaybreaks
\begin{align}
\mathbb{A}_{(\infty)}^1 &=\frac{\Theta _q (x_1x_2)}{\Theta _q (-\sqrt{q}x_1)}\frac{\Theta _q (-\sqrt{q}x_2^{-1}t)}{\Theta _q (-\sqrt{q}x_2^{-1}x_1^{-1}t)}\, \mathcal{J} \left(x_1x_2,qx_1x_2^{-1};qx_1^2;-\sqrt{q}t^{-1};q\right) \nn
&=\Omega ^1_{(\infty)}\mathcal{J} ^1_{(\infty)}\,; \label{blocksnearinfty1} \\
\mathbb{A}_{(\infty)}^2 &= \frac{\Theta _q(x_1^{-1}x_2)}{\Theta _q\big(-\sqrt{q} x_1\big)} \frac{\Theta_q\big(-\sqrt{q} x_2^{-1}t\big)}{\Theta_q \big(-\sqrt{q} x_1x_2^{-1}t\big)}\, \mathcal{J} \left(qx_1^{-1}x_2^{-1} ,x_1^{-1}x_2;qx_1^{-2};-\sqrt{q}t^{-1};q\right) \nn
&=\Omega ^2_{(\infty)}\mathcal{J}^2_{(\infty)}\,, \label{blocksnearinfty2}
\end{align}
\endgroup
where we use an abbreviated notation in terms of $Ω$ and $\J$ to denote the $Θ_q$-prefactors and $\J(⋯)$-function, respectively. The special function $\mathcal{J}(a,b;c;z;q)$ is defined in \eqref{definitionofj} in terms of ${}_2φ_1$ $q$-hypergeometric function. Although we work in a particular $|q|<1$ or $|q|>1$ chamber while solving the LOIs given in \eqref{LOIsqed2}, the solutions we obtain are well-defined and unique in both $q$-chambers.

Naively, the region of validity of the holomorphic blocks $\mathbb{A}^{1,2}_{(\infty)}$ seems to be $|t^{-1}|<1⇒\Re(T)=ζ>0$ for all values of $x_1$ and $x_2$. However, we have to be careful with the series expansion of the $\J$-function around $x_{1,2}→∞$ or $0$. Let us do this exercise for the holomorphic block $\mathbb{A}^1_{(\infty)}$ by using the definition of the $\mathcal{J}$-function \eqref{definitionofj}:
\equ{\mathcal{J}^1_{(∞)}=\frac{(qx_1^{2};q)_\infty}{(x_1x_2,qx_1x_2^{-1};q)_\infty}\sum _{n \geq 0}\frac{(x_1x_2,qx_1x_2^{-1};q)_n}{(qx_1^{2},q;q)_n}\big(-\sqrt{q}t^{-1}\big)^n\,.
}
The finite $q$-Pochhammer symbol with $x_2$ is divergent when $x_2→∞$ and the one with $x_2^{-1}$ is divergent when $x_2→0$. There are no divergences when $x_1→0,∞$. In general, these divergences can be taken care by using the inversion identity given in \eqref{q-pochhammer}. For example, inverting $(x_1x_2;q)_n$ above leads to
\equ{\mathcal{J}^1_{(∞)}=\frac{(qx_1^{2};q)_\infty}{(x_1x_2,qx_1x_2^{-1};q)_\infty}\sum _{n \geq 0}\frac{q^{\frac{n}{2}(n-1)}(q^{-1}x_1^{-1}x_2^{-1};q^{-1})_n (qx_1x_2^{-1};q)_n}{(qx_1^{2},q;q)_n}\big(\sqrt{q}x_1x_2t^{-1}\big)^n\,.
\label{J1infinv}}
Thus, the region of validity near $t,x_2→∞$ is $|x_1x_2t^{-1}|<1⇒m_1+m_2-ζ<0$, where $\Re(X_1)=m_1$ and $\Re(X_2)=m_2$. Similarly, we get $m_1-m_2-ζ<0$ near $t→∞,x_2→0$, so $\J^1_{(∞)}$ is valid in the region given by $m_1+|m_2|-ζ<0$. Repeating the same analysis for $\J^2_{(∞)}$, we get its region as $-m_1+|m_2|-ζ<0$. Thus, the region where both the $\bA_{(∞)}$ blocks can simultaneously exist is the intersection of their respective regions, which can be written concisely as follows:
\equ{\R_{(∞)}=|m_1|+|m_2|-ζ<0\,.
\label{RoVinf}}

The holomorphic blocks near $t=0$ are given by
\begingroup
\allowdisplaybreaks
\begin{align}
\mathbb{A}^1_{(0)}&=\frac{\Theta_q (x_1^{-1}x_2)\Theta _q(x_1x_2)}{\Theta _q(-\sqrt{q}x_1)\Theta _q (x_2^2)}\frac{\Theta _q (-\sqrt{q}x_2^{-1}t)}{\Theta _q (-\sqrt{q}t)}\, \mathcal{J}\big(x_1^{-1}x_2,x_1x_2;x_2^2;-\sqrt{q}t;q \big) \nn
&=\Omega ^1_{(0)} \mathcal{J}^1_{(0)}\,; \label{eq3p2} \\
\mathbb{A}_{(0)}^2 &=\frac{\Theta _q(x_1^{-1}x_2)\Theta _q( x_1x_2)}{\Theta _q(-\sqrt{q}x_1)}\frac{\Theta _q (-\sqrt{q}x_2^{-1}t)}{\Theta _q (-q^{\frac{3}{2}}x_2^{-2}t)}\, \mathcal{J}(qx_1 x_2 ^{-1},qx_1 ^{-1}x_2 ^{-1};q^2 x_2 ^{-2};-\sqrt{q}t;q) \nn
&=\Omega ^2_{(0)} \mathcal{J}^2_{(0)}\,, \label{eq3p3}
\end{align}
\endgroup
Repeating the above analysis for the $\J_{(0)}$ functions, we obtain the following region of validity for the $\mathbb{A}_{(0)}$ blocks:
\equ{\R_{(0)}=|m_1|+|m_2|+ζ<0\,.
\label{RoV0}}

The connection problem for ${}_2φ_1$ $q$-hypergeometric function is quite trivial and is captured by the Watson's connection formula \cite{Watson:1910ghs}. This formula can be adapted to the $\J$-functions (see Appendix \ref{app:FunDefIds} for details) and then the two sets of blocks at $t=0$ and $t=∞$ get connected. For example, we can get for $\J^2_{(0)}$
\begin{equation}
\mathcal{J}^2_{(0)} \rightarrow \frac{\Theta_q(x_1x_2)\Theta_q(-q\sqrt{q}x_1x_2^{-1}t)}{\Theta_q(qx_1^2)\Theta_q(-\sqrt{q}t)}\,\mathcal{J}^1_{(\infty)} +\frac{\Theta_q(x_1^{-1}x_2)\Theta_q(-q\sqrt{q}x_1^{-1}x_2^{-1}t)}{\Theta_q(x_1^2)\Theta_q(-\sqrt{q}t)}\,\mathcal{J}^2_{(\infty)}\,.
\label{connectionfornewblock2I}
\end{equation}
Making use of a similar formula for $\J^1_{(0)}$, we can write down the following connection formula between the holomorphic blocks $\mathbb{A}^{1,2}_{(0)}$ and $\mathbb{A}^{1,2}_{(\infty)}$:
\equ{\begin{pmatrix}
\mathbb{A}^1_{(0)} \\
\mathbb{A}^2_{(0)}
\end{pmatrix}=\begin{pmatrix}
\alpha_q(x_1,x_2,t) & \alpha_q(x_1^{-1},x_2,t) \\
\beta_q(x_1,x_2,t) & \beta_q(x_1^{-1},x_2,t)
\end{pmatrix}
\begin{pmatrix}
\mathbb{A}^1_{(\infty)} \\
\mathbb{A}^2_{(\infty)}
\end{pmatrix},
}
where the matrix elements are given by
\eqs{\alpha_q(x_1,x_2,t) &=\frac{\Theta _q(-\sqrt{q}x_1x_2t)\Theta_q (-\sqrt{q}x_1^{-1}x_2^{-1}t)}{\Theta_q(-\sqrt{q}t)^2}\frac{\Theta_q(x_1^{-1}x_2)^2}{\Theta_q (x_1 ^{-2})\Theta_q (x_2^2)}\,, \\
\beta_q(x_1,x_2,t) &=\frac{\Theta_q(-\sqrt{q}x_1^{-1}x_2^{-1}t)\Theta_q(-q^{\frac{3}{2}}x_1x_2^{-1}t)}{\Theta_q(-\sqrt{q}t)\Theta_q(-q^{\frac{3}{2}}x_2^{-2}t)}\frac{\Theta_q (x_1x_2)\Theta_q(x_1^{-1}x_2)}{\Theta_q (x_1^{-2})}\,·
\label{ellipticbeta}}
Both the $α_q$ and $β_q$-factors are elliptic with respect to all the $q$-difference operators, i.e., $p_{\bullet}α_q=α_q$ and $p_{\bullet}β_q=β_q$.

We now focus on the former set of holomorphic blocks $\bA^{1,2}_{(∞)}$ and work out how the discrete transformations introduced in \eqref{DTs1n2} act on these blocks. This is keeping in mind our goal of obtaining $\cpone$ blocks.

\section{\texorpdfstring{$\bZ_3$ Symmetry and Stokes Phenomena}{Z₃ Symmetry and Stokes Phenomena}}\label{sec:DiscreteTrans}
We saw in Section \ref{sec:SQED2model} that the supersymmetric parameter space of the SQED${}_2$ model is left invariant by the transformations given in \eqref{DTs1n2}. We now wish to study their action on the holomorphic blocks, which will not be trivial and indeed, we will find that it has an interesting relation with the action of mirror symmetry on the holomorphic blocks of $\cpone$ model.

Before proceeding further, we need to `quantize' the transformations given in \eqref{DTs1n2} such that their action on holomorphic blocks is compatible with the appearance of extra $q$-factors in LOIs \eqref{LOIsqed2} when compared to \eqref{ClLOIsqed2}. The $q$-transformations generating the $\bZ_3$ group read as follows:
\begingroup
\allowdisplaybreaks
\begin{align}
\T_{I} &:\quad x_1 \rightarrow \sqrt{\frac{(-\sqrt{q})t}{x_1x_2}}\,,\quad x_2 \rightarrow  \sqrt{(-\sqrt{q})x_1x_2 t}\,,\quad t \rightarrow \frac{x_2}{(-\sqrt{q}) x_1}\,; \label{qTI} \\
(\mathcal{T}_I)^2\equiv \T_{II} &:\quad x_1 \rightarrow  \sqrt{\frac{x_2}{(-\sqrt{q})x_1t}}\,,\quad x_2 \rightarrow  \sqrt{\frac{(-\sqrt{q})x_2 t}{x_1}}\,,\quad t \rightarrow \frac{x_1x_2}{(-\sqrt{q})}\,·
\label{qTII}
\end{align}
\endgroup
It is then clear that the action of these transformations on the holomorphic blocks $\mathbb{A}^{1,2}_{(\infty)}$ will produce a set of new holomorphic blocks. Using the Heine's transformations \cite{Heine:1847,Heine:1878} (denoted by [H]) and Watson's connection formula (denoted by [W]) for the $\J$-functions with $ab=c$ (see Appendix \ref{app:FunDefIds} for more details),
\eqs{\text{[H] :} &\quad \mathcal{J}(a,b;ab;z;q)=\mathcal{J}(z,b;bz;a;q)=\mathcal{J}(z,a;az;b;q)\,; \label{heineforj} \\
\text{[W] :} &\quad \mathcal{J}(a,b;ab;z;q)=\frac{\Theta_q(b)}{\Theta_q(a^{-1}b)}\frac{\Theta_q(az)}{\Theta_q(z)}\mathcal{J}\left(a,\frac{q}{b};\frac{aq}{b};\frac{q}{z};q\right) + (a \leftrightarrow b)\,,
\label{watsonforj}}
we will show that these new blocks are related linearly to the original ones by the following matrix relation:
\eqsa{\mathbb{A}^{1,2}_{(\text{New})}&=\mathcal{M}_{<}\mathbb{A}^{1,2}_{(\infty)} \qquad \text{for $|q|<1$} \\
\mathbb{A}^{1,2}_{(\text{New})}&=\mathcal{M}_{>}\mathbb{A}^{1,2}_{(\infty)} \qquad \text{for $|q|>1$}\,.
}
The matrices $\mathcal{M}_{<}$ and $\mathcal{M}_{>}$ will be triangular and obey the following condition
\equ{\mathcal{M}_{<}\mathcal{M}_{>}^{\sT}=1\,,
}
which suggests that these holomorphic blocks exhibit Stokes-like phenomenon. The above relations then also guarantee that all the partition functions of SQED${}_2$ model constructed out of these holomorphic blocks are invariant under the $\bZ_3$ symmetry discussed above.

\subsection[\texorpdfstring{Transformation $\mathcal{T}_I$}{Transformation T(I)}]{Transformation $\bm{\mathcal{T}_I}$}
Let us first consider the transformation $\T_I$ given in \eqref{qTI}. Acting with this transformation on $\mathcal{J}^{1,2}_{(\infty)}$, we get the following pair of $\mathcal{J}$-functions:
\eqsa{\mathcal{J}^1_{(\infty)} &\xrightarrow{\mathcal{T}_I} \mathcal{J}(qx_1^{-1}x_2^{-1}, -\sqrt{q}t; -q^{3/2}x_1^{-1}x_2^{-1}t; qx_1x_2^{-1}; q) ≡\J^1_I\,, \\
\mathcal{J}^2_{(\infty)} &\xrightarrow{\mathcal{T}_I} 
\mathcal{J}(x_1x_2, -\sqrt{q}t^{-1}; -\sqrt{q}x_1x_2t^{-1}; qx_1x_2^{-1}; q) ≡\J^2_I\,.
\label{JtransTI}}
Analyzing the series expansion of these $\J_I$-functions, we find that their region of validity is
\equ{\R_{I}=m_1-m_2+|ζ|<0 \cap m_1<0 \cap m_2>0\,.
\label{RoVI}}

Using [H] \eqref{heineforj} and [W] \eqref{watsonforj}, we now show that the new pair of $\mathcal{J}$-functions are related to $\mathcal{J}^{1,2}_{(\infty)}$ linearly. This will allow us to define new holomorphic blocks. The following analysis depends on the $q$-chamber. We first focus on the $|q|<1$ chamber.

\subsubsection[\texorpdfstring{For $|q|<1$}{For |q|<1}]{For $\bm{|q|<1}$} \label{sectoriforq<1}
Let us start with the first transformation in \eqref{JtransTI} and apply [H] to get
\equ{\J^1_{(∞)} \xrightarrow{\mathcal{T}_I} \J^1_I \xrightarrow{[H]} \mathcal{J}^2_{(0)}\,.
}
But we know from Section \ref{sec:HoloBlocks} that the $\mathcal{J}^2_{(0)}$ function is related to the $\mathcal{J}^{1,2}_{(\infty)}$ functions by Watson's formula \eqref{connectionfornewblock2I}. Furthermore, the appearance of $\mathcal{J}^2_{(0)}$ suggests we multiply $\J^1_I$ by $\Omega ^2_{(0)}$ to turn it into a new holomorphic block, which after applying [H] and [W] gives:
\equ{\mathbb{A}^1_{I}:=\Omega ^2_{(0)} \J^1_I \longrightarrow β_q(x_1,x_2,t)\big(\bA^2_{(∞)} -\bA^1_{(∞)}\big)\,,
\label{connectionformulaforA1I}}
where the elliptic prefactor $β_q$ is given in \eqref{ellipticbeta}. Since the holomorphic blocks are defined modulo such elliptic functions, we can drop $\beta_q$ in the connection formula for the block $\mathbb{A}^1_I$:
\equ{\mathbb{A}^1_{I} \longrightarrow \mathbb{A}^2_{(\infty)} - \mathbb{A}^1_{(\infty)}\,.
}

Let us now work on the second transformation in \eqref{JtransTI} and again apply [H] to get
\equ{\J^2_{(∞)} \xrightarrow{\T_I} \J^2_I \xrightarrow{[H]} \mathcal{J}_{(\infty)}^1\,.
}
This suggests we can simply define a new holomorphic block as follows
\equ{\mathbb{A}^2_{I} :=\Omega^1_{(\infty)} \J^2_I \longrightarrow \bA^1_{(∞)}\,.
}
This might seem crazy as we have just renamed the $\bA^1_{(∞)}$ block and claimed it is a new block. The reason to do that is simple: despite having an identical series expansion as the holomorphic block $\mathbb{A}^1_{(\infty)}$ in $|q|<1$ chamber, $\mathbb{A}^2_I$ behaves completely differently in $|q|>1$ chamber as we shall see in the next section. Hence as a well-defined function for all values of $q$, we should treat $\mathbb{A}^2_I$ as a different analytic function that satisfies different identities compared to $\mathbb{A}^1_{(\infty)}$.

Thus, we have found the connection matrix for the blocks $\mathbb{A}^{1,2}_I$ and $\mathbb{A}^{1,2}_{(\infty)}$ in $|q|<1$ chamber:
\equ{\begin{pmatrix}
\mathbb{A}^1_{I} \\
\mathbb{A}^2_{I}
\end{pmatrix} \longrightarrow \M^I_{<}
\begin{pmatrix}
\mathbb{A}^1_{(\infty)} \\
\mathbb{A}^2_{(\infty)}
\end{pmatrix};\quad \M^I_{<}=\begin{pmatrix}
-1 & 1 \\
1 & 0
\end{pmatrix}.
\label{stokesfort1withq<1}
}
Let us now find the connection matrix in the other $q$-chamber.

\subsubsection[\texorpdfstring{For $|q|>1$}{For |q|>1}]{For $\bm{|q|>1}$}\label{sec:TIqg1}
The holomorphic blocks $\mathbb{A}^{1,2}_{(\infty)}$ are well-defined in both $|q|\gtrless 1$ chambers. But the Heine's transformation and Watson's formula we used for the $\mathcal{J}$-function are valid only in $|q|<1$ chamber. So we use the property \eqref{propertyofjforbothq} to write both these blocks in terms of $p=q^{-1}$ with $|p|<1$.
\begingroup
\allowdisplaybreaks
\begin{align}
\mathbb{A}_{(\infty)}^1 &=\frac{\Theta _p (px_1^{-1}x_2)\Theta _p (-\sqrt{p}x_1)\Theta _p(-\sqrt{p}x_1^{-1}x_2^{-1}t)}{(p;p)_\infty\Theta _p (-\sqrt{p}x_2^{-1}t)\Theta _p (x_1^2)}\, \mathcal{J} \left(x_1^{-1}x_2^{-1},px_1^{-1}x_2;px_1^{-2};-\sqrt{p}t^{-1};p\right) \nn
&={\Omega'}^1_{(\infty )}{\mathcal{J}'}^1_{(\infty )}\,; \label{blocksnearinfty1p} \\ \mathbb{A}_{(\infty)}^2 &=\frac{\Theta _p(px_1x_2)\Theta _p(-\sqrt{p}x_1)\Theta _p(-\sqrt{p}x_1x_2^{-1}t)}{(p;p)_\infty \Theta _p(px_1^2)\Theta _p(-\sqrt{p}x_2^{-1}t)}\, \mathcal{J}\left(px_1x_2,x_1x_2^{-1};px_1^{2};-\sqrt{p}t^{-1};p\right) \nn
&={\Omega'}^2_{(\infty )}{\mathcal{J}'}^2_{(\infty )}\,. \label{blocksnearinfty2p}
\end{align}
\endgroup

Now we focus on the holomorphic block $\mathbb{A}^1_I$ and rewrite it as a $p$-hypergeometric function by using \eqref{propertyofjforbothq} as follows:
\eqsg{\mathbb{A}^1_I={\Omega}^1_I\mathcal{J}\(-\sqrt{p}t^{-1},px_1x_2;-p\sqrt{p} x_1x_2t^{-1};x_1x_2^{-1};p\), \\
{\Omega}^1_I=\frac{\Theta _p(-\sqrt{p}t)\Theta _p(-\sqrt{p}x_1)\Theta _p(-p\sqrt{p}x_2^{2}t^{-1})}{(p;p)_\infty \Theta _p(x_1x_2^{-1})\Theta _p(-\sqrt{p}x_2^{-1}t)\Theta _p(-p\sqrt{p}x_1x_2t^{-1})}\,·
}
Applying only [H] now, we get
\equ{\mathbb{A}^1_I \longrightarrow \mathbb{A}^2_{(\infty)}\,,
}
up to an elliptic factor given by
\[\frac{{\Omega}^1_I}{\Omega^2_{(\infty)}} =\frac{\Theta _p(-\sqrt{p}t)\Theta _p(-p\sqrt{p}x_2^{2}t^{-1})\Theta _p(px_1^2)}{ \Theta _p(x_1x_2^{-1})\Theta _p(px_1x_2)\Theta _p(-\sqrt{p}x_1x_2^{-1}t)\Theta _p(-p\sqrt{p}x_1x_2t^{-1})}\,·\]
This ratio of theta factors is equivalent to the elliptic factor $\beta_q$ (with $p=q^{-1}$) appearing in equation \eqref{connectionformulaforA1I} in the analysis of $|q|<1$ chamber.

Using the same property \eqref{propertyofjforbothq}, we can write the other new block $\mathbb{A}^2_I$ as follows:
\equ{\mathbb{A}^2_I=\frac{\Theta _p(-\sqrt{p}x_1)\Theta _p(-\sqrt{p}t)}{(p;p)_\infty \Theta _p(-\sqrt{p}x_2^{-1}t)}\, \mathcal{J}\(-\sqrt{p}t,x_1^{-1}x_2^{-1};-\sqrt{p}x_1^{-1}x_2^{-1}t;x_1x_2^{-1};p\).
}
In order to relate this block to $\mathbb{A}^{1,2}_{(\infty)}$ given in \eqref{blocksnearinfty1p} and \eqref{blocksnearinfty2p}, we apply a combined [H] and [W] to get
\equ{\bA^2_I  \longrightarrow \bA^1_{(∞)} +\bA^2_{(∞)}\,.
}

Thus, we have found that the blocks $\mathbb{A}^{1,2}_I$ are also related to the $\mathbb{A}^{1,2}_{(\infty)}$ blocks in the $|q|>1$ chamber and the connection matrix reads as follows:
\equ{\begin{pmatrix}
\mathbb{A}^1_{I} \\
\mathbb{A}^2_{I}
\end{pmatrix} \longrightarrow \M^I_{>}
\begin{pmatrix}
\mathbb{A}^1_{(\infty)} \\
\mathbb{A}^2_{(\infty)}
\end{pmatrix};\quad \M^I_{>}=
\begin{pmatrix}
0 & 1 \\
1 & 1
\end{pmatrix}.
\label{stokesfort1withq>1}
} 
The connection matrices in \eqref{stokesfort1withq<1} and \eqref{stokesfort1withq>1} satisfy the condition $\mathcal{M}^{I}_{<}\big(\mathcal{M}^{I}_{>}\big)^{\sT}=1$ associated with Stokes matrices as promised.

\subsection[\texorpdfstring{Transformation $\mathcal{T}_{II}$}{Transformation T(II)}]{Transformation $\bm{\mathcal{T}_{II}}$}
We now consider the transformation $\mathcal{T}_{II}≡(\T_I)^2$ given in \eqref{qTII}. Acting with this transformation on $\mathcal{J}^{1,2}_{(\infty)}$, we get a new pair of $\mathcal{J}$-functions
\eqsa{\mathcal{J}^1_{(\infty)} & \xrightarrow{\mathcal{T}_{II}} \mathcal{J}\(x_1^{-1}x_2, -\sqrt{q}t^{-1}; -\sqrt{q}x_1^{-1}x_2t^{-1}; qx_1^{-1}x_2^{-1}; q\) ≡\J^1_{II}\,, \\
\mathcal{J}^2_{(\infty)} &\xrightarrow{\mathcal{T}_{II}} \mathcal{J}\(qx_1x_2^{-1}, -\sqrt{q}t; -q\sqrt{q}x_1x_2^{-1}t; qx_1^{-1}x_2^{-1}; q\) ≡\J^2_{II}\,.
\label{JtransTII}}
Again, the analysis of the series expansion of these $\J_{II}$-functions leads to their region of validity:
\equ{\R_{II}=-m_1-m_2+|ζ|<0 \cap m_1>0 \cap m_2>0\,.
\label{RoVII}}

Following similar steps discussed for $\T_I$ in the previous section, we show that this new pair of $\mathcal{J}$-functions gives rise to new holomorphic blocks of the SQED${}_2$ model, which are also related to the $\mathbb{A}^{1,2}_{(\infty)}$ blocks by Stokes phenomenon. Let us first carry out this analysis in the $|q|<1$ chamber.

\subsubsection[\texorpdfstring{For $|q|<1$}{For |q|<1}]{For $\bm{|q|<1}$}
We start with the first transformation in \eqref{JtransTII} and apply [H] to get

\equ{\J^1_{(∞)} \xrightarrow{\T_{II}} \J^1_{II} \xrightarrow{[H]} \mathcal{J}^2_{(\infty)}\,.
}

\noindent Following the same line of argument used in section \ref{sectoriforq<1}, we define a new block as follows

\equ{{\mathbb{A}}^1_{II}:=\Omega^2_{(\infty)}\J^1_{II} \longrightarrow \bA^2_{(∞)}\,.
}

Next, we work on the second transformation in \eqref{JtransTII} and again apply [H] to get

\equ{\J^2_{(∞)} \xrightarrow{\T_{II}} \J^2_{II} \xrightarrow{[H]} \mathcal{J}^2_{(0)}\,.
}

\noindent This chain is again similar to what we saw in section \ref{sectoriforq<1} and we can use the same prefactor $\Omega^2_{(0)}$ and proceed. But we can also proceed slightly differently here so as not to have any elliptic factors in the final connection formula. For that, let us define

\equ{\bA^2_{II}:=\wh{Ω}^2_{(0)}\J^2_{II}\,,
}

\noindent where $\wh{Ω}^2_{(0)}$ differs from $\Omega^2_{(0)}$ by an elliptic factor and it explicitly reads

\[\wh{Ω}^2_{(0)}=-\frac{\Theta _q(qx_1^2)\Theta _q(-\sqrt{q}t)\Theta _q (-\sqrt{q}x_2^{-1}t)}{\Theta _q(-q\sqrt{q}x_1x_2^{-1}t)\Theta _q (-\sqrt{q}x_1) \Theta _q (-\sqrt{q}x_2^{-1}x_1^{-1}t)}\,·
\]

\noindent Now with the action of [H] and [W], the connection formula for $\bA^2_{II}$ reads

\equ{\bA^2_{II} \longrightarrow \bA^2_{(∞)} -\bA^1_{(∞)}\,.
}

Thus, the connection matrix for the blocks $A^{1,2}_{II}$ and $A^{1,2}_{(∞)}$ in $|q|<1$ chamber is as follows:
\equ{\begin{pmatrix}
\mathbb{A}^1_{II} \\
\mathbb{A}^2_{II}
\end{pmatrix} \longrightarrow \M^{II}_{<}
\begin{pmatrix}
\mathbb{A}^1_{(\infty)} \\
\mathbb{A}^2_{(\infty)}
\end{pmatrix};\quad \M^{II}_{<}=
\begin{pmatrix}
0 & 1 \\
-1 & 1
\end{pmatrix}.
\label{stokesfort2withq<1}}
Let us now work in the other $q$-chamber.

\subsubsection[\texorpdfstring{For $|q|>1$}{For |q|>1}]{For $\bm{|q|>1}$}
We now turn to $|q|>1$ chamber and perform a similar analysis as done in section \ref{sec:TIqg1}. We express the newly defined blocks $\mathbb{A}^{1,2}_{II}$ in terms of $p=q^{-1}$ using \eqref{propertyofjforbothq} as follows:
\begin{align}
\mathbb{A}^1_{II} &=\frac{\Theta _p(-\sqrt{p}x_1)\Theta _p(-\sqrt{p}t)}{(p;p)_\infty \Theta _p(-\sqrt{p}x_2^{-1}t)}\, \mathcal{J}\(-\sqrt{p}t, x_1 x_2^{-1}; -\sqrt{p}x_1x_2^{-1}t; x_1^{-1}x_2^{-1}; p\); \\
\mathbb{A}^2_{II} &=Ω^2_{II}\mathcal{J}\(-\sqrt{p}t^{-1},px_1^{-1}x_2;-p\sqrt{p}x_1^{-1}x_2t^{-1};x_1^{-1}x_2^{-1};p\), \\
\text{with }Ω^2_{II} &=-\frac{\Theta _p(-\sqrt{p}x_1)\Theta _p(x_1x_2^{-1})\Theta _p(-\sqrt{p}x_1x_2t^{-1})}{(p;p)_\infty\Theta _p(x_1^2)\Theta _p(-\sqrt{p}x_2^{-1}t)}\,·\nonumber
\end{align}
Applying a combined [H] and [W] to $\bA^1_{II}$, we get
\begin{align}
\mathbb{A}^1_{II} \longrightarrow \mathbb{A}^1_{(\infty)}+\mathbb{A}^2_{(\infty)}\,;
\end{align}
and applying only [H] to $\bA^2_{II}$, we get
\begin{align}
\mathbb{A}^2_{II} \longrightarrow -\mathbb{A}^1_{(\infty)}\,.
\end{align}

Thus, the connection matrix between the holomorphic blocks $\mathbb{A}^{1,2}_{II}$ and $\mathbb{A}^{1,2}_{(\infty)}$ in $|q|>1$ chamber is given by:
\equ{\begin{pmatrix}
\mathbb{A}^1_{II} \\
\mathbb{A}^2_{II}
\end{pmatrix} \longrightarrow \M^{II}_{>}
\begin{pmatrix}
\mathbb{A}^1_{(\infty)} \\
\mathbb{A}^2_{(\infty)}
\end{pmatrix};\quad \M^{II}_{>}=
\begin{pmatrix}
1 & 1 \\
-1 & 0
\end{pmatrix}.
\label{stokesfort2withq>1}}
The connection matrices in \eqref{stokesfort2withq<1} and \eqref{stokesfort2withq>1} satisfy the condition $\mathcal{M}^{II}_{<}\big(\mathcal{M}^{II}_{>}\big)^{\sT}=1$ as before. This again suggests Stokes phenomenon at work.

\subsection{Regions of Validity}
We have found four pairs of blocks for the SQED${}_2$ model and their respective regions of validity as given in \eqref{RoVinf}, \eqref{RoV0}, \eqref{RoVI} and \eqref{RoVII}. A plot of the boundaries of these regions is shown in Figure \ref{fig:SRegs}. We see that the whole parameter space gets divided into six distinct, contiguous regions. The four pairs of blocks we have found form a complete basis of blocks in four of those six regions. This suggests the existence of two more pairs of blocks, let's call them $\bA^{1,2}_{III,IV}$, following from the action of the $\bZ_3$ symmetry on $\bA^{1,2}_{(0)}$ blocks, as one might expect. Indeed, such blocks do exist with the `correct' regions of validity. It turns out that these two sets of blocks can also be obtained by acting with another $\bZ_3$ symmetry on $\bA^{1,2}_{(∞)}$ directly, i.e., by introducing two `new' transformations $\T_{III},\T_{IV}$ (given explicitly in \eqref{qTIII}, \eqref{qTIV} and discussed further in Appendix \ref{app:ExtraDTs}) obtained by inverting $t$ in $\T_{I},\T_{II}$. Thus, we find six pairs of bona fide holomorphic blocks for SQED${}_2$ model to go along with the six regions (the remaining two regions are given in \eqref{RoVIII} and \eqref{RoVIV}) uniquely covering the whole parameter space of the SQED${}_2$ model. We collect these six regions here for quick reference:
\eqsn{\R_{(∞)} &=|m_1|+|m_2|-ζ<0\,, \\
\R_{(0)} &=|m_1|+|m_2|+ζ<0\,, \\
\R_{I} &=m_1-m_2+|ζ|<0 \cap m_1<0 \cap m_2>0\,, \\
\R_{II} &=-m_1-m_2+|ζ|<0 \cap m_1>0 \cap m_2>0\,, \\
\R_{III} &=-m_1+m_2+|ζ|<0 \cap m_1>0 \cap m_2<0\,, \\
\R_{IV} &=m_1+m_2+|ζ|<0 \cap m_1<0 \cap m_2<0\,.
}
\fig{!h}{\includegraphics[scale=0.55]{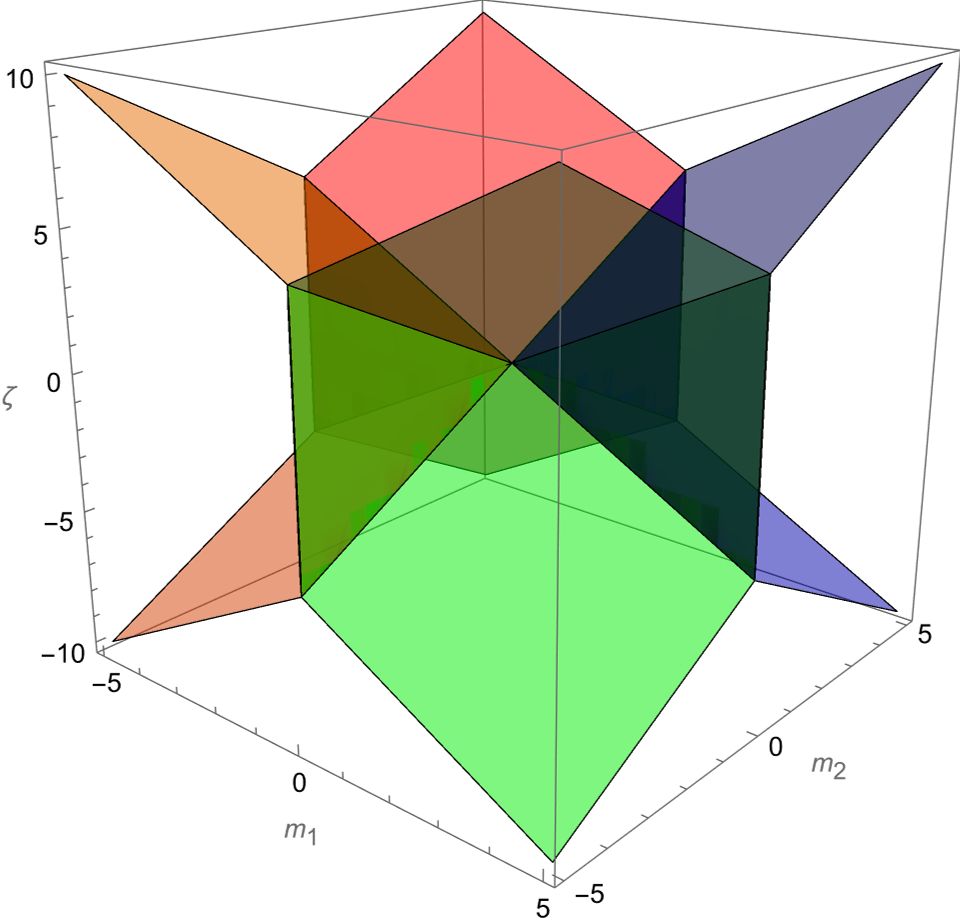}
\caption{The regions of validity for the six pairs of holomorphic blocks of the SQED${}_2$ model. The top and bottom regions ($\R_{(∞)}$ and $\R_{(0)}$) host $\bA^{1,2}_{(∞)}$ and $\bA^{1,2}_{(0)}$ blocks, respectively. The red (rear-facing) quadrant ($\R_{I}$) hosts the $\bA^{1,2}_{I}$ blocks, blue (right-facing) one ($\R_{II}$) hosts $\bA^{1,2}_{II}$, green (front-facing) one ($\R_{III}$) hosts $\bA^{1,2}_{III}$, and orange (left-facing) one ($\R_{IV}$) hosts $\bA^{1,2}_{IV}$.}
\label{fig:SRegs}}

This is quite reminiscent of the Stokes phenomenon, though some connection matrices (as those discussed in the Appendix \ref{app:ExtraDTs}) have some entries which are elliptic factors and not integers, i.e., $M^α{}_β\notin GL(N,\bZ)$. Nonetheless, there still exist nontrivial connection matrices between all these 6 pairs of holomorphic blocks and analytic continuation is required to move across these regions in the full parameter space, just like in Stokes phenomenon. We depict the relationship between these six pairs of blocks\footnote{Strictly speaking, only the $\J$-functions since the complete blocks are obtained after multiplying with appropriate $Ω$'s and acting with [H] or [H]+[W] transformations.} in Figure \ref{fig:AllHBs}.
\fig{!h}{\tdplotsetmaincoords{80}{125}
\begin{tikzpicture}[tdplot_main_coords,scale=3]
\node (inf) at (0,0,1) {$\forclr{red}{blue}{darkgreen}{orange}{\J^{1,2}_{(∞)}}{-0.3}{0}$};
\node (Z) at (0,0,-1) {$\forclr{orange}{darkgreen}{blue}{red}{\J^{1,2}_{(0)}}{-0.26}{0}$};
\node (I) at (-1,-1,0) {${\color{red!25}\J^{1,2}_{I}}$};
\node (II) at (-1,1,0) {${\color{blue}\J^{1,2}_{II}}$};
\node (III) at (1,1,0) {${\color{darkgreen}\J^{1,2}_{III}}$};
\node (IV) at (1,-1,0) {${\color{orange}\J^{1,2}_{IV}}$};
\path[<->,>=stealth,thick,font=\footnotesize]
(I) edge[red!50!blue!25] node[style={fill=white}]{$\U$} (II)
(IV) edge[orange!50!red!25] node[style={fill=white}]{$\V$} (I);
\path[->,>=stealth,thick,font=\footnotesize]
(inf) edge[red!25] node[style={fill=white}]{$\T_{I}$} (I)
(inf) edge[blue] node[style={fill=white}]{$\quad\T_{II}$} (II)
(inf) edge[darkgreen] node[style={fill=white}]{$\;\;\T_{III}\quad$} (III)
(inf) edge[orange] node[style={fill=white}]{$\T_{IV}$} (IV);
\path[->,>=stealth,thick,font=\footnotesize]
(Z) edge[red!25] node[style={fill=white}]{$\T_{III}$} (I)
(Z) edge[blue] node[style={fill=white}]{$\quad\T_{IV}$} (II)
(Z) edge[darkgreen] node[style={fill=white}]{$\T_{I}\;\;$} (III)
(Z) edge[orange] node[style={fill=white}]{$\;\T_{II}$} (IV);
\path[<->,>=stealth,thick,font=\footnotesize]
(II) edge[blue!50!darkgreen] node[style={fill=white}]{$\V$} (III)
(III) edge[darkgreen!50!orange] node[style={fill=white}]{$\U$} (IV);
\end{tikzpicture}
\caption{The six pairs of $\J$-functions of the SQED${}_2$ model. The colour coding and positioning (at the vertices of an octahedron) of the $\J$-functions reflects the associated region of validity as depicted in Figure \ref{fig:SRegs}. Also, note two inversions, $\U: x_1 \rightarrow x_1^{-1}$ and $\V: x_2 \rightarrow qx_2^{-1}$.}
\label{fig:AllHBs}}

\newpage
Having seen the Stokes phenomena in the SQED${}_2$ model, let us now relate it to the well-known Stokes phenomena of the $\cpone$ model.

\section{\texorpdfstring{From SQED$\bm{{}_2}$ to $\bm{\cpone}$ Model}{From SQED₂ to CP¹ Model}}\label{sec:DecLim}
Let us recall that the $\mathbb{CP}^1$ model \cite{Beem:2012mb, Ashok:2019gee} is described by a $U(1)$ gauged linear sigma model with two chiral fields having same charge under the gauge group. This GLSM flows to a non-linear sigma model with $\mathbb{CP}^1$ target space in the infrared. The theory has flavour symmetry $SU(2)\times U(1)_J$ which is broken to $U(1)_V \times U(1)_J$ when the (complexified) twisted mass associated to $U(1)_V$ is turned on. The chiral fields are not charged under the $U(1)_R$ symmetry.
\equ{T^{\cpone}[\vec{\phi}]=\left\{\vphantom{\pmat{ \\[1.75cm] }}\right.\quad
\begin{array}{c|cc}Q & \phi_1 & \phi_2 \\ \hline
G & 1 & 1 \\
V & 1 & -1 \\
J & 0 & 0 \\
R & 0 & 0
\end{array}\qquad
\begin{array}{c|cccc}k & G & V & J & R \\ \hline
G & 0 & 0 & 1 & 0 \\
V & 0 & 0 & 0 & 0 \\
J & 1 & 0 & 0 & 0 \\
R & 0 & 0 & 0 & \star
\end{array}
\label{cp1modeldef}}

The line operator identities of the $\mathbb{CP}^1$ model are
\eqsa{\wh{p}_y +(\wh{y}^{-1} -\wh{x} -\wh{x}^{-1})+\wh{p}_y^{-1} &\simeq 0 \\
q^{-\frac{1}{2}}\wh{p}_x\wh{p}_y -\wh{x}\big(q^{\frac{1}{2}}\wh{p}_x +\wh{p}_y\big) +1 &\simeq 0\,.
\label{LOIcp1}}
The first LOI involving only the $q$-difference operator $\widehat{p}_y$ determines the $q$-hypergeo-metric series of the holomorphic block while the second LOI fixes the prefactors involving only the twisted flavour mass $x$. The singularity structure of the first LOI is such that both regular singular point ($y=\infty$) and irregular singular point ($y=0$) exist. Following the procedure reviewed in \cite{TAB}, one obtains the following holomorphic blocks as a solution near the regular singular point \cite{Ashok:2019gee}
\begin{align}
\mathbb{B}^1_I &=\frac{\Theta _q(y)}{\Theta _q(-\sqrt{q}x)\Theta _q(x^{-1}y)}\,\mathcal{J}(xy^{-1},x^2;q)\,; \label{B1ICP1def}\\
\mathbb{B}^2_I &=\frac{\Theta _q(y)}{\Theta _q(-\sqrt{q}x)\Theta _q(xy)}\,\mathcal{J}(x^{-1}y^{-1},x^{-2};q) \,.
\label{B2ICP1def}
\end{align}
Here, $\mathcal{J}(x,y;q)$ is the so-called Hahn-Exton $q$-Bessel function, which is expressed in terms of the $q$-hypergeometric function ${}_1\phi_1$ as follows
\begin{equation}
\mathcal{J}(x,y;q)=(qy;q)_\infty\,{}_1\phi_1(0;qy;q,qx)\,.
\end{equation}

The solutions near the irregular singular point $y=0$ are hard to find as they diverge in general. But in this case, one can use $q$-Borel resummation and express the solutions in terms of the solutions found near the regular singular point \cite{OHY2016}. Using this technique, it is possible (though arduous) to find two more pairs of holomorphic blocks \cite{Ashok:2019gee}. These holomorphic blocks can also be found by using the $\mathbb{Z}_3$ symmetry transformations, better known as the \emph{mirror symmetry} transformations \cite{Beem:2012mb}:
\equ{\begin{aligned}
\omega &:\qquad x \rightarrow \sqrt{x^{-1}y}\,,\qquad y \rightarrow \sqrt{x^{-3}y^{-1}}\,; \\
\omega^2 &:\qquad x \rightarrow \sqrt{x^{-1}y^{-1}}\,,\qquad y \rightarrow \sqrt{x^3y^{-1}}\,.
\end{aligned}
\label{MirrorCP1}}
Of course, $ω^3=1$. The supersymmetric parameter space of $\mathbb{CP}^1$ model remains invariant under these transformations. By applying these transformations, one can get two new pairs of holomorphic blocks, which are related to $\mathbb{B}^{1,2}_I$ via Stokes phenomena. The first pair of new blocks reads
\begin{align}
\mathbb{B}^1_{II}&=\frac{\Theta _q(y)}{\Theta _q(-\sqrt{q}x)\Theta _q(x^{-1}y)}\,\mathcal{J}(x^2,xy^{-1};q)\,; \\
\mathbb{B}^2_{II}&=\frac{\Theta _q(qx^2)\Theta _q(y)}{\Theta _q(-\sqrt{q}x)\Theta_q(qxy)\Theta _q(x^{-1}y)}\,\mathcal{J}(xy,x^{-1}y;q)\,,
\end{align}
and the second pair similarly reads
\begin{align}
\mathbb{B}^1_{III}&=\frac{\Theta _q(y)}{\Theta _q(-\sqrt{q}x)\Theta _q(xy)}\,\mathcal{J}(x^{-2},x^{-1}y^{-1};q)\,; \\
\mathbb{B}^2_{III}&=\frac{\Theta _q(x^2)\Theta _q(y)}{\Theta _q(-\sqrt{q}x)\Theta _q(xy)\Theta _q(qx^{-1}y)}\,\mathcal{J}(x^{-1}y,xy;q)\,.
\end{align}

From the description of SQED${}_2$ model in \eqref{sqed2model}, it is clear that the limit $m_2→±∞$ effectively decouples the pair of chiral fields $(φ_3,φ_4)$ charged under $F_2$ and reduces the model to that of $\cpone$ given in \eqref{cp1modeldef}.\footnote{We will not consider the limit $m_1→±∞$ that decouples the $(φ_1,φ_2)$ pair as it does not lead to the `canonical' $\cpone$ model of \eqref{cp1modeldef} due to the nonzero R-charge of $\phi _4$.} This means that these chiral multiplets contribute neither to the partition functions nor to the holomorphic blocks. Let us now see what this decoupling limit does to the SQED${}_2$ holomorphic blocks we found in the previous sections.

We start with the holomorphic blocks found by solving the LOIs near $t=\infty$ in Section \ref{sec:HoloBlocks} and focus on the $m_2→∞ ⇒ x_2\rightarrow \infty$ limit.\footnote{\label{fn12}The other limit $m_2→-∞ ⇒ x_2→0$ also leads to a nice decoupling limit but for the trio of $\bA_{(∞)}$, $\bA_{III}$ and $\bA_{IV}$ blocks. However, these SQED${}_2$ blocks reduce to $\cpone$ blocks with extra elliptic factors so we do not discuss this case explicitly.} Expanding the holomorphic block $\mathbb{A}^1_{(\infty)}$ by using the $\mathcal{J}$-function in \eqref{J1infinv}, we get
\eqst{\mathbb{A}^1_{(\infty)}=\frac{\Theta _q\big(-\sqrt{q}x_2^{-1}t\big)(qx_1^{2},qx_1^{-1}x_2^{-1},q;q)_\infty}{\Theta _q\big(-\sqrt{q} x_1\big)\Theta _q \big(-\sqrt{q}x_1^{-1}x_2^{-1}t\big)(qx_1x_2^{-1};q)_\infty} \\
×\sum _{n \geq 0}\frac{(-1)^nq^{\frac{n}{2}(n-1)}(q^{-1}x_1^{-1}x_2^{-1};q^{-1})_n (qx_1x_2^{-1};q)_n}{(qx_1^{2},q;q)_n}\big(-\sqrt{q}x_1x_2t^{-1}\big)^n\,.
\label{eq5p11}}
The $q$-Pochhammer symbols (both finite and infinite) with $x_2^{-1}$ become unity in the limit $x_2 \rightarrow \infty$ so we do not worry about them. It is also evident that we need to send the FI parameter $t$ to $\infty$ as well in order to get a well-defined FI parameter for the $\mathbb{CP}^1$ model. Strictly speaking, this double scaling limit is what we mean by the decoupling limit of SQED${}_2$ model. To identify the resulting expression with the holomorphic blocks of $\mathbb{CP}^1$ model, we are then led to the following parameter identifications
\equ{-\sqrt{q}x_2^{-1}t = y\,, \qquad x_1 = x \qquad \text{as }\; x_2,t \rightarrow \infty\,.
\label{identificationI}}
Thus \eqref{eq5p11} reduces to
\eqs{\lim_{x_2,t \rightarrow \infty} \mathbb{A}^1_{(\infty)} &=\frac{\Theta _q\big(y \big)}{\Theta _q\big(-\sqrt{q} x\big)\Theta _q \big(x^{-1}y \big)} (qx^{2},q;q)_\infty \sum _{n \geq 0}\frac{(-1)^nq^{\frac{n}{2}(n-1)}}{(qx^{2},q;q)_n}\big(qxy^{-1}\big)^n \nn
&=\mathbb{B}^1_I\,,
}
where $\bB^1_I$ is given in \eqref{B1ICP1def}. Following similar steps (with $x→x^{-1}$), we recover the second holomorphic block $\mathbb{B}^2_I$ \eqref{B2ICP1def} of the $\mathbb{CP}^1$ model from $\mathbb{A}^2_{(\infty)}$ block of the SQED${}_2$ model, i.e.,
\equ{\lim_{x_2,t \rightarrow \infty} \mathbb{A}^2_{(\infty)} =\mathbb{B}^2_I\,.
}

This decoupling limit can also be taken for the holomorphic blocks $\mathbb{A}^{1,2}_I$ and $\mathbb{A}^{1,2}_{II}$ to get the other two pairs of the $\cpone$ holomorphic blocks. This analysis is quite straightforward, so we skip the details and just summarize the result as follows:
\[\begin{tabular}{c@{$\qquad\longrightarrow\qquad$}c}
SQED${}_2$ &  $\mathbb{CP}^1$ \\ 
\hline 
$\vphantom{\Big(}\mathbb{A}^{1,2}_{(\infty)}$ & $\mathbb{B}^{1,2}_I$ \\[1mm]
$\mathbb{A}^{1,2}_{I}$ & $\mathbb{B}^{2,1}_{II}$ \\[1mm]
$\mathbb{A}^{1,2}_{II}$ & $\mathbb{B}^{1,2}_{III}$
\end{tabular}\]
Note that not only the holomorphic blocks but also the LOIs \eqref{LOIsqed2} and the discriminant locus \eqref{spmD} of SQED${}_2$ reduce correctly to those of $\cpone$ in this decoupling limit. We leave the proof of the former to the reader but show the latter in Figure \ref{fig:sqed2cp1shift}. Moreover, the transformations $\T_{I,II}$ \eqref{qTI}-\eqref{qTII} also reduce exactly to the mirror symmetry transformations $ω,ω^2$ given in \eqref{MirrorCP1}, respectively.
\fig{!h}{$\vc{\begin{tikzpicture}[scale=1.75]
\node (t) at (0,1) {$t =-\frac{x_2+x_2^{-1}±2}{x_1+x_1^{-1}∓2}$};
\node (y) at (0,-1) {$-x_2^{-1}t =y =\frac{1}{x+x^{-1}±2}$};
\path[->,>=stealth,thick,font=\footnotesize]
(t) edge node[label={[shift={(0.2,-1)}]{\rotatebox{-90}{$x_2,t→∞$}}}]{} (y);
\end{tikzpicture}} \qquad \vc{\includegraphics[scale=0.45]{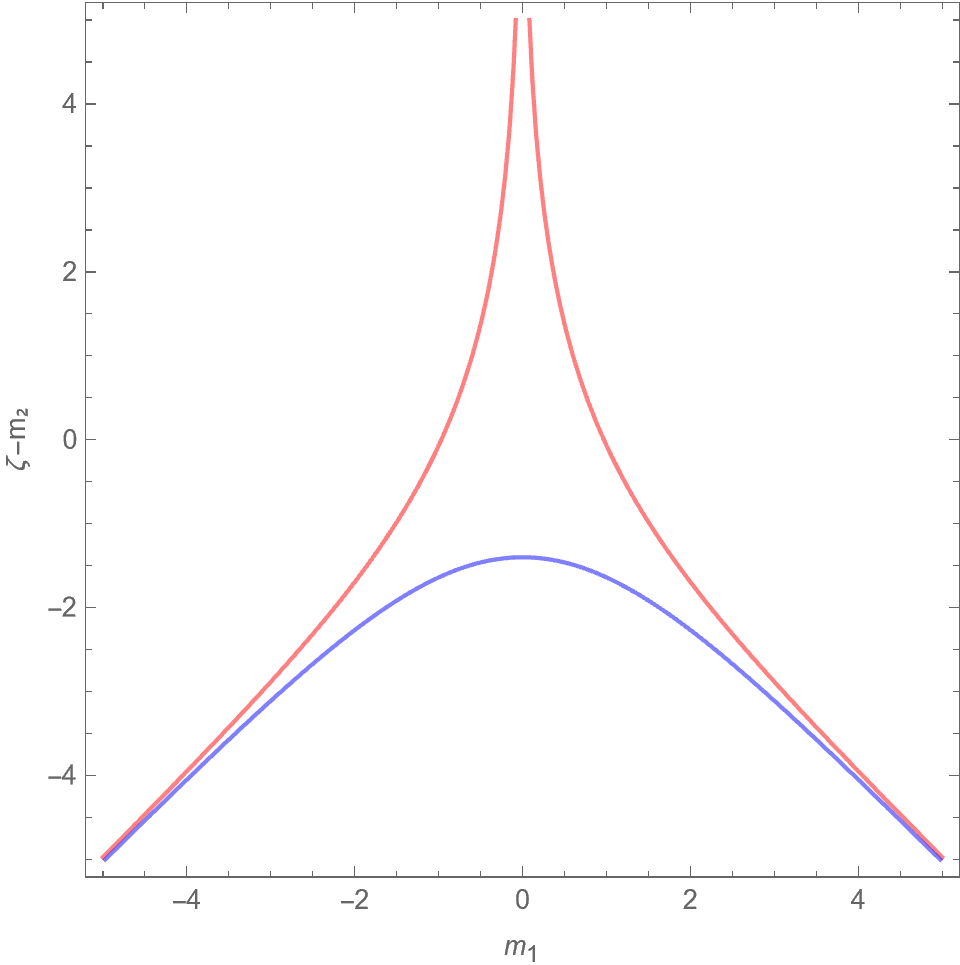}}$
\caption{Discriminant locus of $\cpone$ as obtained in the decoupling limit of that of SQED${}_2$. (For plotting purposes, $m_2=5$ is chosen to represent $x_2→∞$.)}
\label{fig:sqed2cp1shift}}

\subsection{The Stokes Regions and Matrices}
Now we look at the Stokes regions and find that the correct $\cpone$ Stokes regions \cite{Beem:2012mb,Ashok:2019gee} are obtained in the decoupling limit. The limiting procedure picks out three regions from the top-rear region (upper half of the red-blue quadrants) of Figure \ref{fig:SRegs} and is shown in Figure \ref{fig:SRegscp1}.
\fig{!h}{$\vc{\includegraphics[scale=0.45]{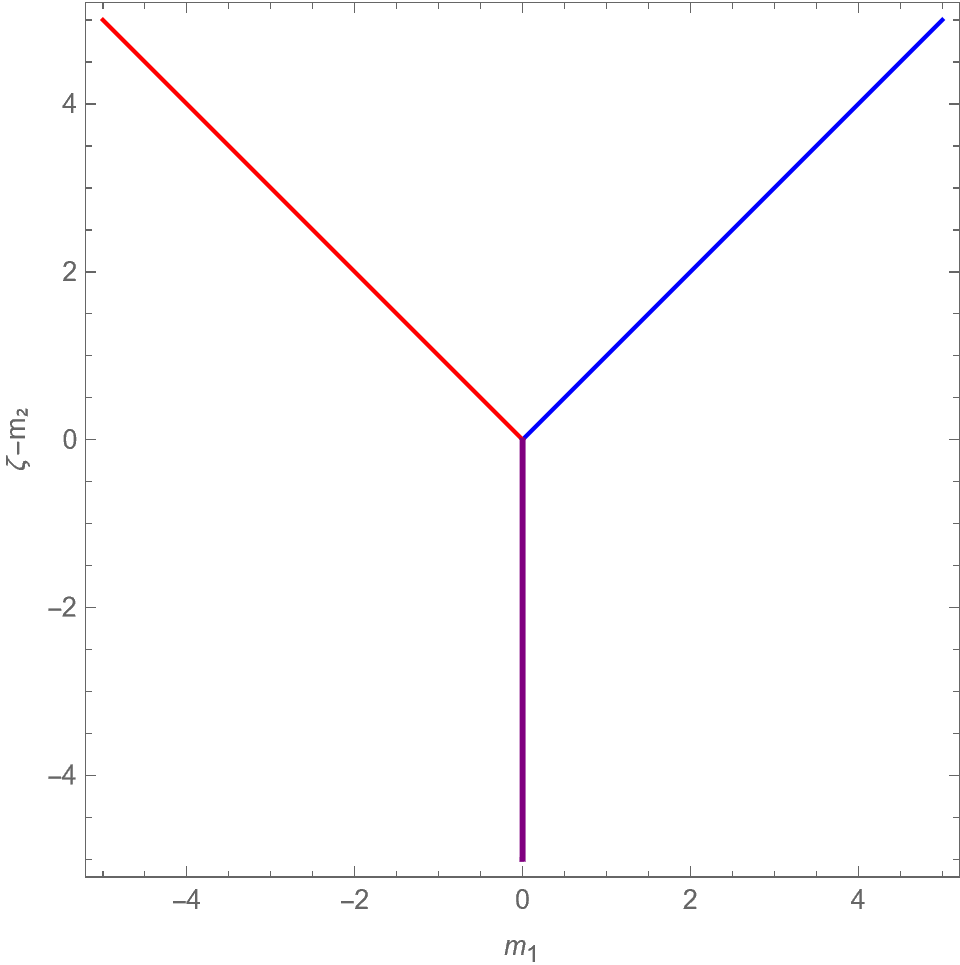}} \qquad \vc{\begin{tikzpicture}[scale=1]
\node (BI) at (0,3) {$\twoclr{red}{blue}{\bB^{1,2}_{I}}{0.12}$};
\node (BII) at (-3,0) {$\color{red}\bB^{1,2}_{II}$};
\node (BIII) at (3,0) {$\color{blue}\bB^{1,2}_{III}$};
\path[->,>=stealth,thick,font=\footnotesize]
(BI) edge[red] node[style={fill=white}]{$ω$} (BII)
(BI) edge[blue] node[style={fill=white}]{$ω^2$} (BIII);
\path[<->,>=stealth,thick,font=\footnotesize]
(BII) edge[red!50!blue] node[style={fill=white}]{$\U$} (BIII);
\end{tikzpicture}}$
\caption{Left: The three Stokes regions of the $\cpone$ model as obtained in the decoupling limit of SQED${}_2$ model (again, $m_2=5$ represents $x_2→∞$ as in Figure \ref{fig:sqed2cp1shift}). Right: The associated holomorphic blocks and the relevant transformations, which act only on the $\J$-functions appearing inside $\bB$'s. (The complete blocks are then obtained by multiplying $\J$'s with appropriate prefactors, as in the case of the SQED${}_2$ blocks.)}
\label{fig:SRegscp1}}

We can now also reproduce the Stokes matrices between the holomorphic blocks of the $\mathbb{CP}^1$ model from the connection matrices of the SQED${}_2$ model derived in the previous section. For example, the Stokes matrices between the holomorphic blocks $\mathbb{B}^{1,2}_I$ and $\mathbb{B}^{1,2}_{II}$ can be extracted from \eqref{stokesfort1withq<1} and \eqref{stokesfort1withq>1} as follows:
\[\begin{array}{c@{\qquad}|@{\qquad}c}
\begin{tikzcd}[row sep=huge, column sep = normal]
\begin{pmatrix}
\mathbb{A}^1_{I} \\
\mathbb{A}^2_{I}
\end{pmatrix} \arrow{r}{\T_I} \arrow[swap]{d}{\rotatebox{-90}{$x_2,t \rightarrow \infty$}} & \M^I_{<}\begin{pmatrix}
\mathbb{A}^1_{(\infty)} \\
\mathbb{A}^2_{(\infty)}
\end{pmatrix} \arrow{d}{\rotatebox{-90}{$x_2,t \rightarrow \infty$}} \\
\begin{pmatrix}
\mathbb{B}^2_{II} \\
\mathbb{B}^1_{II}
\end{pmatrix} \arrow[swap]{r}{ω} & \M^I_{<}\begin{pmatrix}
\mathbb{B}^1_{I} \\
\mathbb{B}^2_{I}
\end{pmatrix}
\end{tikzcd} & \begin{tikzcd}[row sep=huge, column sep = normal]
\begin{pmatrix}
\mathbb{A}^1_{I} \\
\mathbb{A}^2_{I}
\end{pmatrix} \arrow{r}{\T_I} \arrow[swap]{d}{\rotatebox{-90}{$x_2,t \rightarrow \infty$}} & \M^I_{>}\begin{pmatrix}
\mathbb{A}^1_{(\infty)} \\
\mathbb{A}^2_{(\infty)}
\end{pmatrix} \arrow{d}{\rotatebox{-90}{$x_2,t \rightarrow \infty$}} \\
\begin{pmatrix}
\mathbb{B}^2_{II} \\
\mathbb{B}^1_{II}
\end{pmatrix}  \arrow[swap]{r}{ω} & \M^I_{>}\begin{pmatrix}
\mathbb{B}^1_{I} \\
\mathbb{B}^2_{I}
\end{pmatrix}
\end{tikzcd} \\
\M^I_{<}=\begin{pmatrix}
-1 & 1 \\
1 & 0
\end{pmatrix} & \M^I_{>}=\begin{pmatrix}
0 & 1 \\
1 & 1
\end{pmatrix}
\end{array}\]
Similarly, we get the following Stokes matrices between the holomorphic blocks $\mathbb{B}^{1,2}_I$ and $\mathbb{B}^{1,2}_{III}$ of the $\mathbb{CP}^1$ model from \eqref{stokesfort2withq<1} and \eqref{stokesfort2withq>1}:
\[\begin{array}{c@{\qquad}|@{\qquad}c}
\begin{tikzcd}[row sep=huge, column sep = normal]
\begin{pmatrix}
\mathbb{A}^1_{II} \\
\mathbb{A}^2_{II}
\end{pmatrix} \arrow{r}{\T_{II}} \arrow[swap]{d}{\rotatebox{-90}{$x_2,t \rightarrow \infty$}} & \M^{II}_{<}\begin{pmatrix}
\mathbb{A}^1_{(\infty)} \\
\mathbb{A}^2_{(\infty)}
\end{pmatrix} \arrow{d}{\rotatebox{-90}{$x_2,t \rightarrow \infty$}} \\
\begin{pmatrix}
\mathbb{B}^1_{III} \\
\mathbb{B}^2_{III}
\end{pmatrix}  \arrow[swap]{r}{ω^2} & \M^{II}_{<}\begin{pmatrix}
\mathbb{B}^1_{I} \\
\mathbb{B}^2_{I}
\end{pmatrix}
\end{tikzcd} & \begin{tikzcd}[row sep=huge, column sep = normal]
\begin{pmatrix}
\mathbb{A}^1_{II} \\
\mathbb{A}^2_{II}
\end{pmatrix} \arrow{r}{\T_{II}} \arrow[swap]{d}{\rotatebox{-90}{$x_2,t \rightarrow \infty$}} & \M^{II}_{>}\begin{pmatrix}
\mathbb{A}^1_{(\infty)} \\
\mathbb{A}^2_{(\infty)}
\end{pmatrix} \arrow{d}{\rotatebox{-90}{$x_2,t \rightarrow \infty$}} \\
\begin{pmatrix}
\mathbb{B}^1_{III} \\
\mathbb{B}^2_{III}
\end{pmatrix}  \arrow[swap]{r}{ω^2} & \M^{II}_{>}\begin{pmatrix}
\mathbb{B}^1_{I} \\
\mathbb{B}^2_{I}
\end{pmatrix}
\end{tikzcd} \\
\M^{II}_{<}=\begin{pmatrix}
0 & 1 \\
-1 & 1
\end{pmatrix} & \M^{II}_{>}=\begin{pmatrix}
1 &1 \\
-1 & 0
\end{pmatrix}
\end{array}\]
Thus, we have correctly reproduced the Stokes matrices $\M$'s of $\cpone$ as derived in \cite{Beem:2012mb, Ashok:2019gee} in the decoupling limit \eqref{identificationI} of SQED${}_2$ discussed above.

\section{Summary and Outlook}\label{sec:Outlook}
In this note, we have studied the Stokes phenomena exhibited by the holomorphic blocks of SQED${}_2$ and $\cpone$ models. First, we solved the LOIs of SQED${}_2$ model to get two pairs of holomorphic blocks $\bA^{1,2}_{(∞)}$ and $\bA^{1,2}_{(0)}$. Second, we identified a $\bZ_3$ symmetry generated by a transformation $\T_{I}$ that leaves the supersymmetric parameter space of the SQED${}_2$ model invariant. Under the action of the transformations $\T_I$ and $\T_{II}≡(\T_{I})^2$, we found two new pairs of holomorphic blocks for the SQED${}_2$ model that are related to $\bA^{1,2}_{(∞)}$ blocks by Stokes phenomenon. Finally, we reduced this triplet of SQED${}_2$ holomorphic blocks to $\cpone$ blocks in the decoupling limit given by \eqref{identificationI} reproducing the Stokes matrices of the latter model exactly.

As mentioned in the Introduction, extracting the Stokes behaviour of $\cpone$ blocks from that of SQED${}_2$ model was our primary motivation. But we also analyzed the connection problem of holomorphic blocks for the SQED${}_2$ model itself. That is, in addition to the trio of $\bA^{1,2}_{(∞)}$, $\bA^{1,2}_{I}$ and $\bA^{1,2}_{II}$ blocks, we also have another trio of $\bA^{1,2}_{(0)}$, $\bA^{1,2}_{III}$ and $\bA^{1,2}_{IV}$ blocks related by the same $\bZ_3$ symmetry. We found that the regions of validity for these six pairs of holomorphic blocks are non-overlapping and uniquely cover the whole parameter space of the SQED${}_2$ model, as shown in figures \ref{fig:SRegs} and \ref{fig:AllHBs}. This is also how we had found the Stokes regions for the $\cpone$ model using the algebraic approach in \cite{Ashok:2019gee}. However, it turns out that the latter trio of blocks do not admit any nice decoupling limit to the $\cpone$ blocks. That is, even though we have six pairs of bona fide holomorphic blocks for SQED${}_2$ model, only half of them admit a nice decoupling limit to $\cpone$ holomorphic blocks. The failure of these holomorphic blocks to have any nice limit to $\cpone$ blocks can be attributed to the $q$-hypergeometric function ${}_2\phi _1(a,b;c;q,z)$ appearing in these blocks either a) diverging, e.g. $\bA_{III,IV}$ in $x_2→∞$ limit, or b) failing to produce a ${}_1φ_1$-function required for a $\cpone$ block, e.g. $\bA_{(0)}$ in $x_2→∞$ limit. Similar behaviour is exhibited in the $x_2→0$ limit by these six blocks, but with the role reversal of $\bA_{I,II}$ and $\bA_{III,IV}$. Let us reiterate that we do not consider the $m_1→±∞$ limit in this note due to our choice of R-charge for the chiral field $\phi_4$ in \eqref{sqed2model}, which breaks the symmetry between the pairs of chiral fields $(φ_1,φ_2)$ and $(φ_3,φ_4)$ such that decoupling the former pair does not lead to the `canonical' $\cpone$ model \eqref{cp1modeldef}. Thus, although the $\J$-functions appearing in Figure \ref{fig:AllHBs} reduce to the relevant $\J$-functions of the $\cpone$ model, the corresponding $Ω$-prefactors will not reduce to that of the $\cpone$ model, due to crucial relative sign differences between the third LOI of \eqref{LOIsqed2} and the second LOI of \eqref{LOIcp1}.

Let us now discuss the implications for $\cpnnn$ and SQED$_{N+1}$ holomorphic blocks in general. On the one hand, the blocks for $\mathbb{CP}^N$ model are given by ${}_N\phi_N(\vec{0};a_{1,⋯,N};q,y)$ $q$-hypergeometric functions so in analogy to $\cpone$ \cite{Ashok:2019gee}, it may not be too far-fetched to expect that the Stokes phenomena is encoded in the connection problem of these $q$-hypergeometric functions with $N \geq 2$. As discussed in the Introduction, this is precisely due to the presence of an irregular singularity of the $q$-difference equation (at $y=0$) satisfied by these $q$-hypergeometric functions. Thus, it would be interesting to solve the connection problem for the ${}_N\phi_N(⋯)$-functions with $N \geq 2$ to directly study the Stokes phenomena for $\mathbb{CP}^N$. On the other hand, the blocks of SQED$_{N+1}$ models are given by the ${}_{N+1}\phi_N (a_{1,\cdots,N+1};b_{1,\cdots,N};q,t)$ $q$-hypergeometric functions \cite{Nieri:2015yia} and their connection problem is again captured by appropriate Watson's formula. However, the Stokes-like phenomena exhibited by the holomorphic blocks of SQED${}_{N+1}$ is not encoded in this connection formula as we saw in the case of the SQED${}_2$ model. For SQED${}_2$, we are required to use the Heine's transformation that identifies two $\mathcal{J}$-functions with different parameters, in order to find the new blocks $\mathbb{A}_{I,II}$ that are analytic continuation of the $\mathbb{A}_{(\infty)}$ blocks. The higher-$N$ analogue of Heine's transformation for ${}_{N+1}\phi_N(\cdots)$-function is not so obvious and might not even lead to the same $q$-hypergeometric function. For example, the holomorphic blocks of the SQED${}_3$ model are expressed in terms of ${}_3\phi_2(a_1,a_2,a_3;b_1,b_2;q,t)$ function for which the Heine-like transformation leads to the $q$-Lauricella function \cite{Andrews1972,KAJIHARA200453,Bhatnagar2019}. However, we think it should be possible to find connection matrices acting on an appropriate basis consisting of these functions and then studying a particular decoupling limit to access the Stokes phenomenon for $\cpnnn$ model should be straightforward. We hope to address these $N≥2$ issues in the near future.

\section*{Acknowledgements}
We thank Sujay Ashok and Dileep Jatkar for collaboration at the initial stages of this work. We also acknowledge insightful discussions with them and their helpful comments on the draft of this note.

\appendix
\section{Special Functions: Definitions and Identities}\label{app:FunDefIds}
\begin{itemize}
\item $q$-Pochhammer symbol (finite case):
\begin{equation}
(x;q)_n =\prod _{i=0}^{n-1}(1-q^i x)  \,.
\end{equation}
An inversion identity is useful to know,
\equ{(x;q)_n =(-x)^nq^{\frac{n(n-1)}{2}}(q^{-1}x^{-1};q^{-1})_n\,.
\label{q-pochhammer}}
Another identity to reduce $q$-factors in the argument also comes in handy,
\begin{equation}
(q^k x;q)_n=\frac{(q^nx;q)_k}{(x;q)_k}(x;q)_n\,.
\end{equation}

\item $q$-Pochhammer symbol (infinite case):
\begin{align}
 (x;q)_\infty  &= \begin{cases}
 \prod _{n=0}^\infty (1-xq^n) & \text{for } |q|<1 \\
 \prod _{n=1}^\infty (1-xq^{-n})^{-1} & \text{for } |q|>1 \end{cases} \\
 &=\sum _{n=0}^\infty \frac{(-1)^n q^{\frac{1}{2}n(n-1)}}{(q;q)_n} x^n \,.
\label{qPochinfinite}
\end{align}
A couple of useful identities are:
\begin{align}
(x;q)_\infty &=\frac{1}{(q^{-1}x;q^{-1})_{\infty}}\,, \\
(q^kx;q)_\infty &=\frac{(x;q)_\infty}{(x;q)_k}\,·
\end{align}

\item $q$-Jacobi theta function: 
\begin{align}
\Theta_q (x) &= (x;q)_\infty (qx^{-1};q)_\infty (q;q)_\infty \equiv (x,qx^{-1},q;q)_\infty\,,  \label{qthetafunction} 
\end{align}
with $x \in \mathbb{C} $  for $|q|<1$ and $x \in \mathbb{C} \setminus q^{\mathbb{Z}}$ for $|q|>1$. Using Jacobi triple product formula, $q$-Jacobi theta function can be shown to have the following series expansion:
\begin{equation}
\Theta _q (x)= \begin{cases} \sum _{n \in \mathbb{Z}}(-1)^n q^{\frac{n}{2}(n-1)}x^n & \text{for } |q|<1 \\
\left(\sum _{n \in \mathbb{Z}}(-1)^n q^{-\frac{n}{2}(n+1)}x^n\right)^{-1} & \text{for } |q|>1\,. \end{cases}
\label{qthetafunction1}
\end{equation}
Some useful identities for this functions are (valid in both $|q|\gtrless 1$ chambers):
\begin{align}
\Theta_q(x) &=\Theta _q(qx^{-1})\,, \label{thetaproperty} \\
\Theta_q(qx)&=(-x^{-1})\Theta _q(x)\,, \\
\Theta_q(x)&=\frac{1}{\Theta _{q^{-1}}(x^{-1})}\,·
\end{align}

\item $q$-hypergeometric functions have the following power series expansion for $|x|<1$ and $|q| \gtrless 1$:
\begin{align}
{}_r\phi_s(a_1,...,a_r; b_1,..,b_s; q, x) =\sum _{n\geq 0}\frac{(a_1,...,a_r;q)_n}{(b_1,..,b_s ;q)_n (q;q)_n}\left[(-1)^n q^{\frac{n}{2}(n-1)}\right]^{1+s-r} x^n\,,
\end{align}
where $0\leq s\leq r $. The relevant function for SQED${}_2$ is ${}_2φ_1(a,b;c;q,z)$ so we enumerate here some of its most important identities. One set of identities is given by the three Heine's transformation formulae (in $|q|<1$ chamber):
\begin{align}
{}_2\phi_1(a,b;c;q,z) &=\frac{(b,az;q)_\infty}{(c,z;q)_\infty}\, {}_2\phi_1\(\frac{c}{b},z;az;q,b\) \label{hein1} \\
&=\frac{\(\frac{c}{b},bz;q\)_\infty}{(c,z;q)_\infty}\, {}_2\phi_1\(\frac{abz}{c},b;bz;q,\frac{c}{b}\) \label{hein2} \\
&=\frac{\big(\frac{abz}{c};q\big)_\infty}{(z;q)_\infty}\, {}_2\phi_1\(\frac{c}{a},\frac{c}{b};c;q,\frac{abz}{c}\). \label{hein3} 
\end{align}
Another identity is the Watson's formula: 
\eqst{{}_2 \phi _1 (a,b;c;q,z)=\frac{\(b,\frac{c}{a};q\)_\infty }{\big(c,\frac{b}{a};q\big)_\infty }\frac{\Theta _q(az)}{\Theta _q(z)} {}_2 \phi _1 \left(a,\frac{aq}{c};\frac{aq}{b};q,\frac{cq}{abz}\right) \\ 
+\frac{\(a,\frac{c}{b};q\)_\infty}{\(c,\frac{a}{b};q\)_\infty}\frac{\Theta _q(bz)}{\Theta _q(z)}{}_2 \phi _1 \left(b,\frac{bq}{c};\frac{bq}{a};q,\frac{cq}{abz}\right).
}

\item The $\mathcal{J}(a,b;c;z;q)$ function is defined in terms of ${}_2\phi_1$ $q$-hypergeometric function as follows:
\equ{\mathcal{J}(a,b;c;z;q):=\frac{(c;q)_\infty}{(a,b;q)_\infty}\,{}_2\phi_1(a,b;c;q;z) \qquad \text{for both } |q| \gtrless 1\,.
\label{definitionofj}}
The following symmetry property of $\mathcal{J}(a,b;c;z;q)$ function under exchange of $a$ and $b$ then follows from the definition in both $|q|\gtrless 1$ chamber
\equ{\mathcal{J}(a,b;c;z;q)=\mathcal{J}(b,a;c;z;q)\,. \label{symmetryofj}
}
The inversion property of $\mathcal{J}(a,b;ab;z;q)$ with $p=q^{-1}$ is also valid in both $|q| \gtrless 1$ chamber:
\equ{\frac{\Theta _p(pa)\Theta _p(pb)}{ (p;p)_\infty\Theta _p(pab)} \mathcal{J}(a^{-1},b^{-1};a^{-1}b^{-1};pz;p)=\mathcal{J}(a,b;ab;z;q)\,.
\label{propertyofjforbothq}}
The parameters of the $\mathcal{J}$-functions appearing in the expressions of holomorphic blocks of SQED${}_2$ model satisfy the condition $ab=c$. In this special case, the Heine's transformations simplify to
\equ{\mathcal{J}(a,b;ab;z;q)=\mathcal{J}(z,b;bz;a;q)=\mathcal{J}(z,a;az;b;q) \,.
}
 
The Watson's formula for ${}_2\phi_1$ function given above can also be translated straightforwardly to the $\J$-function. We just give the relation for the case $ab=c$ as this is the only case used in the main text:
\equ{\mathcal{J}(a,b;ab;z;q)=\frac{\Theta_q(b)}{\Theta_q\big(\frac{b}{a}\big)}\frac{\Theta_q(az)}{\Theta_q(z)}\mathcal{J}\left(a,\frac{q}{b};\frac{aq}{b};\frac{q}{z};q\right) + (a \leftrightarrow b)\,.
}

\end{itemize}

\section{Solving the LOIs}\label{app:DeriveHBs}
In this appendix, we present the derivation of the holomorphic blocks $\mathbb{A}^{1,2}_{(\infty)}$ from the LOIs \eqref{LOIsqed2} of the SQED${}_2$ model in detail. These LOIs take the form of the so-called $q$-difference equations, once we identify the operator $\widehat{p}_{x}$ with the $q$-difference operator $\sigma_q(x)=q^{x\frac{d}{dx}}$. Their action on $\widehat{x}$ is as follows
\equ{\widehat{p}\,\widehat{x}=q\widehat{x}\,\widehat{p}\,.
}
We shall focus first on the $q$-difference equation involving only the difference operator $\widehat{p}_t$ (first LOI in \eqref{LOIsqed2}) and write its solution in the following form
\equ{\mathbb{A}(x_1,x_2,t;q)=f(x_1,x_2;q)g(x_1,x_2,t;q)\,.
}
The $t$-independent factor will be determined by the second and third LOIs of \eqref{LOIsqed2}. The $q$-difference equation satisfied by $g(x_1,x_2,t;q)$ then is the following
\equ{\left[(1-q^{-1}\widehat{x}_2\widehat{p}_t)(1-\widehat{x}_2^{-1}\widehat{p}_t)+\sqrt{q}\widehat{t} (1-\widehat{x}_1 \widehat{p}_t)(1-\widehat{x}_1^{-1}\widehat{p}_t)\right]g =0\,.
\label{differenceeqnpy}}
This is a second order $q$-difference equation, so there exist two linearly independent solutions for $g(x_1,x_2,t;q)$. The singular points of the $q$-difference equation are $t=0$ and $t=\infty$, both of which are regular singular. We are interested in finding the solutions near $t=\infty$, so we introduce the following variable\footnote{This step is not needed while solving \eqref{differenceeqnpy} near $t=0$. Other steps can be followed identically to obtain \eqref{eq3p2}-\eqref{eq3p3}.}
\equ{u=t^{-1}\,·
}
In terms of $u$, we express the difference operator $\widehat{p}_t \equiv \sigma _q(t)=\sigma _q^{-1}(u)$. From now on, we use $\s_q$ notation for the $\wh{p}$ operators and omit the hats over all the variables to simplify notation. The $q$-difference equation \eqref{differenceeqnpy} then takes the following form
\equ{\left[\sqrt{q}u(1-qx_2^{-1}\sigma _q(u))(1-x_2\sigma _q(u))+(1-x_1^{-1} \sigma _q(u))(1-x_1\sigma _q(u))\right]g =0\,.
}
In order to map the above $q$-difference equation to the standard $q$-Goursat form (see \cite{Ashok:2019gee} for some references and details), we introduce a new function $h(x_1,x_2,u;q)$ as follows
\equ{g(x_1,x_2,u;q)=\frac{\Theta_q(-\sqrt{q}x_2u)}{\Theta_q(-\sqrt{q}x_1x_2u)}h(x_1,x_2,u;q) \,.
}
The function $h(x_1,x_2,u;q)$ satisfies the following equation
\equ{\left[\sqrt{q}u(1-qx_1x_2^{-1}\sigma _q(u))(1-x_1x_2\sigma _q(u))+(1-\sigma _q(u))(1-x_1^2\sigma _q(u))\right]h =0\,.
}
One of the solutions to the above $q$-difference equation is now straightforward
\equ{h(x_1,x_2,u;q)={}_2\phi _1\left(x_1x_2, qx_1x_2^{-1}; qx_1^2; q, -\sqrt{q}u \right).
}
Thus, the first solution near $t=\infty$ can be expressed as follows
\equ{\mathbb{A}^1_{(\infty)}=f(x_1,x_2;q)\frac{\Theta_q (-\sqrt{q}x_2t^{-1})}{\Theta_q (-\sqrt{q}x_1x_2t^{-1})}\, {}_2\phi_1\left(x_1x_2, qx_1x_2^{-1}; qx_1^2; q, -\sqrt{q}t^{-1}\right).
\label{A1infwithf}}

In order to determine the holomorphic block completely, we have to fix the prefactor $f(x_1,x_2;q)$, which is done by substituting \eqref{A1infwithf} in the two remaining LOIs in \eqref{LOIsqed2}. This analysis is tedious but straightforward, and we write the resulting two equations for $f(x_1,x_2;q)$:
\begingroup
\allowdisplaybreaks
\begin{align}
f(qx_1,x_2;q) &=\left(-\sqrt{q}x_2^{-1}\right) \frac{(1-x_1x_2)(1-qx_1x_2^{-1})}{(1-qx_1^2)(1-q^2x_1^2)} f(x_1,x_2;q)\,; \\
f(x_1,qx_2;q) &=\left(-x_1^{-1}x_2^{-1}\right) \frac{(1-x_1x_2)}{(1-x_1x_2^{-1})}f(x_1,x_2;q)\,.
\end{align} 
\endgroup
These two equations can be solved by choosing (up to elliptic factors)
\equ{f(x_1x_2;q)=\frac{\Theta _q(x_1x_2)}{\Theta _q(-\sqrt{q}x_1)}\frac{(qx_1^2;q)_\infty}{(x_1x_2, qx_1x_2^{-1}; q)_\infty}\,·
}
So the first holomorphic block defined near $t=\infty$ is given by
\begin{align}
\mathbb{A}^1_{(\infty)} &=\frac{\Theta_q(x_1x_2)\Theta_q (-\sqrt{q}x_2t^{-1})}{\Theta_q(-\sqrt{q}x_1)\Theta_q (-\sqrt{q}x_1x_2t^{-1})}\frac{(qx_1^2; q)_\infty}{(x_1x_2, qx_1x_2^{-1}; q)_\infty}\, {}_2\phi_1\!\left(\!x_1x_2, \frac{qx_1}{x_2}; qx_1^2; q, \frac{-\sqrt{q}}{t}\right) \nn[2mm]
&=\frac{\Theta_q(x_1x_2)\Theta_q (-\sqrt{q}x_2t^{-1})}{\Theta_q(-\sqrt{q}x_1)\Theta_q(-\sqrt{q}x_1x_2t^{-1})}\, \mathcal{J}\left(x_1x_2, qx_1x_2^{-1}; qx_1^2; -\sqrt{q}t^{-1}; q\right).
\end{align}
In the last step, we used the definition of $\mathcal{J}$-function given in \eqref{definitionofj}.

In order to find the second solution, one can use the procedure we discussed in \cite{Ashok:2019gee} (for a succinct review, see \cite{TAB}). But we can take a shortcut and write down the second solution by replacing $x_1 \rightarrow x_1^{-1}$ in $\mathbb{A}^1_{(\infty)}$. We are allowed to do this because the first and second LOIs in \eqref{LOIsqed2} are invariant under the $\bZ_2$ symmetry $\U$. So we get the second holomorphic block defined near $t=\infty$ as follows:
\equ{\mathbb{A}^2_{(\infty)} =\frac{\Theta_q(x_1^{-1}x_2)\Theta_q (-\sqrt{q}x_2t^{-1})}{\Theta_q(-\sqrt{q}x_1)\Theta_q(-\sqrt{q}x_1^{-1}x_2t^{-1})}\, \mathcal{J}\left(x_1^{-1}x_2, qx_1^{-1}x_2^{-1}; qx_1^{-2}; -\sqrt{q}t^{-1}; q\right).
}

\section{\texorpdfstring{Another $\bZ_3$ Symmetry}{Another Z₃ Symmetry}}\label{app:ExtraDTs}
In this appendix, we explore another $\mathbb{Z}_3$ symmetry generated by two transformations $\mathcal{T}_{III}$ and $\mathcal{T}_{IV}$ that are obtained from $\T_{I}$ and $\T_{II}$ by inverting $t$:
\eqs{\T_{III} &:\quad x_1 \rightarrow  \sqrt{\frac{(-\sqrt{q})}{x_1x_2 t}}\,,\quad x_2 \rightarrow  \sqrt{\frac{(-\sqrt{q})x_1x_2}{t}}\,,\quad t \rightarrow \frac{(-\sqrt{q})x_1}{ x_2}\,; \label{qTIII}\\
(\T_{III})^2 ≡\T_{IV} &:\quad x_1 \rightarrow \sqrt{\frac{x_2t}{(-\sqrt{q})x_1}}\,,\quad x_2 \rightarrow \sqrt{\frac{(-\sqrt{q})x_2}{x_1t}}\,,\quad t \rightarrow \frac{(-\sqrt{q})}{x_1x_2} \,·
\label{qTIV}}
The holomorphic blocks we get due to the action of these transformations on the blocks defined near $t=\infty$, do not produce the holomorphic blocks of $\mathbb{CP}^1$ model in the decoupling limit \eqref{identificationI}, but do so in another limit as mentioned in footnote \ref{fn12}. Regardless, they are legitimate holomorphic blocks of SQED${}_2$ model since they reproduce correct partition functions for this model and are again connected to the $\mathbb{A}^{1,2}_{(\infty)}$ blocks by Stokes-like phenomenon.

We follow a similar procedure as the one detailed in Section \ref{sec:DiscreteTrans} to find new holomorphic blocks consisting of the new transformed $\mathcal{J}$-functions. Thus, we do not repeat the full analysis but present only the final results here. The general procedure can be summarized as follows: apply only [H] on the $\mathcal{J}$-function with $\(-\sqrt{q}t^{-1}\)$ as its argument; but on the $\mathcal{J}$-function with $\(-\sqrt{q}t\)$ as its argument, apply both [H] and [W]; finally, multiply an appropriate prefactor to turn the $\mathcal{J}$-functions into holomorphic blocks.

\subsection[\texorpdfstring{Transformation $\mathcal{T}_{III}$}{Transformation T(III)}]{Transformation $\bm{\mathcal{T}_{III}}$}
The $\mathcal{J}$-functions of the $A^{1,2}_{(∞)}$ blocks transform under $\mathcal{T}_{III}$ \eqref{qTIII} as follows:
\eqsa{\mathcal{J}^1_{(\infty)} &\xrightarrow{\mathcal{T}_{III}} \mathcal{J}\(qx_1^{-1}x_2^{-1}, -\sqrt{q}t^{-1}; -q\sqrt{q}x_1^{-1}x_2^{-1}t^{-1}; x_1^{-1}x_2; q\) , \\
\mathcal{J}^2_{(\infty)} &\xrightarrow{\mathcal{T}_{III}} 
\mathcal{J}\(x_1x_2, -\sqrt{q}t;-\sqrt{q}x_1x_2t; x_1^{-1}x_2; q\).
}
The above $\J$ functions lead to the following region of validity:
\equ{\R_{III}=-m_1+m_2+|ζ|<0 \cap m_1>0 \cap m_2<0\,.
\label{RoVIII}}

The new holomorphic blocks of the SQED${}_2$ model then read
\begin{align}
\mathbb{A}_{III}^1 &:=\Omega^2_{(\infty)} \mathcal{J}\(qx_1^{-1}x_2^{-1}, -\sqrt{q}t^{-1}; -q\sqrt{q}x_1^{-1}x_2^{-1}t^{-1}; x_1^{-1}x_2; q\); \\
\mathbb{A}_{III}^2 &:=\Omega^2_{III} \mathcal{J}\(x_1x_2, -\sqrt{q}t; -\sqrt{q}x_1x_2t; x_1^{-1}x_2; q\),
\end{align}
where the prefactor $\Omega ^2_{III}$ is equivalent to $\Omega _{(0)}^1$ up to an elliptic factor and is given by
\equ{\Omega^2_{III}=\frac{\Theta_q(qx_1^2)\Theta_q(x_1x_2)\Theta_q(-\sqrt{q}t)\Theta_q(-\sqrt{q}x_2^{-1}t)}{\Theta_q(-\sqrt{q}x_1)\Theta_q(x_1^{-1}x_2)\Theta_q(-\sqrt{q}x_1x_2t^{-1})\Theta_q(-\sqrt{q}x_1x_2t)}\,·
}
The connection matrix relating the above holomorphic blocks to $\mathbb{A}^{1,2}_{(\infty)}$ in the $|q|<1$ chamber then reads
\equ{\begin{pmatrix}
\mathbb{A}^1_{III} \\
\mathbb{A}^2_{III}
\end{pmatrix} = \begin{pmatrix}
0 & 1 \\
1 & \chi
\end{pmatrix} \begin{pmatrix}
\mathbb{A}^1_{(\infty)} \\
{\mathbb{A}^2_{(\infty)}}
\end{pmatrix},
}
where $\chi$ is an elliptic factor given by the following ratio of theta functions
\equ{\chi =\frac{\Theta _q ^2(x_1x_2)\Theta _q (x_1^{-2})}{\Theta _q ^2 (x_1^{-1}x_2)\Theta _q (x_1^{2})} \frac{\Theta _q(-\sqrt{q}x_1^{-1}x_2t)\Theta _q (-\sqrt{q}x_1x_2^{-1}t)}{\Theta _q(-\sqrt{q}x_1x_2t)\Theta _q (-\sqrt{q}x_1^{-1}x_2^{-1}t)}\,·
}
The analysis in the $|q|>1$ chamber can be done similarly to the one presented in Section \ref{sec:DiscreteTrans}. So we skip the details here. The connection matrices relating the holomorphic blocks $\mathbb{A}^{1,2}_{III}$ and $\mathbb{A}^{1,2}_{(\infty)}$ for $|q|<1$ and $|q|>1$ can also be shown to satisfy the consistency condition $\mathcal{M}_{<}\mathcal{M}_{>}^{\sT}=1$.

\subsection[\texorpdfstring{Transformation $\mathcal{T}_{IV}$}{Transformation T(IV)}]{Transformation $\bm{\mathcal{T}_{IV}}$}
The $\mathcal{J}$-functions of the $\mathbb{A}^{1,2}_{(\infty)}$ blocks transform under $\mathcal{T}_{IV}$ \eqref{qTIV} as follows
\eqsa{\mathcal{J}^1_{(\infty)} &\xrightarrow{\mathcal{T}_{IV}} 
\mathcal{J}\(-\sqrt{q}t, x_1^{-1}x_2; -\sqrt{q}x_1^{-1}x_2t; x_1x_2; q\) , \\
\mathcal{J}^2_{(\infty)} &\xrightarrow{\mathcal{T}_{IV}} 
\mathcal{J}\(-\sqrt{q}t^{-1}, qx_1x_2^{-1}; -q\sqrt{q}x_1x_2^{-1}t^{-1}; x_1x_2; q\).
}
The above $\J$ functions lead to the following region of validity:
\equ{\R_{IV}=m_1+m_2+|ζ|<0 \cap m_1<0 \cap m_2<0\,.
\label{RoVIV}}

The new holomorphic blocks then turn out to be the following:
\begin{align}
\mathbb{A}^1_{IV} &:=\Omega^2_{III}\mathcal{J}\(-\sqrt{q}t, x_1^{-1}x_2; -\sqrt{q}x_1^{-1}x_2t; x_1x_2; q\); \\
\mathbb{A}^2_{IV} &:=\Omega_1^{(\infty)}\mathcal{J}\(-\sqrt{q}t^{-1}, qx_1x_2^{-1}; -q\sqrt{q}x_1x_2^{-1}t^{-1}; x_1x_2; q\).
\end{align}
The connection matrix relating the holomorphic blocks $\mathbb{A}^{1,2}_{(\infty)}$ to these new blocks $\mathbb{A}^{1,2}_{IV}$ in the $|q|<1$ chamber then reads
\equ{\begin{pmatrix}
\mathbb{A}^1_{IV} \\
\mathbb{A}^2_{IV}
\end{pmatrix} =\begin{pmatrix}
1 & \chi \\
1 & 0
\end{pmatrix} \begin{pmatrix}
\mathbb{A}^1_{(\infty)} \\
{\mathbb{A}^2_{(\infty)}}
\end{pmatrix}.
}
We again skip the analysis for $|q|>1$ chamber as this can be done easily just by replacing $q=p^{-1}$ with $|p|<1$. The connection matrices relating the holomorphic blocks $\mathbb{A}^{1,2}_{IV}$ and $\mathbb{A}^{1,2}_{(\infty)}$ for $|q|<1$ and $|q|>1$ also satisfy the consistency condition $\mathcal{M}_{<}\mathcal{M}_{>}^{\sT}=1$.

\bibliographystyle{hephys}
\bibliography{3dRefs}

\end{document}